\documentclass[]{relaxed_system_lab}

\usepackage[utf8]{inputenc}
\usepackage[T1]{fontenc}
\usepackage{geometry}
\usepackage{amsmath,amssymb}  %
\usepackage{charter}

\usepackage[toc,page,header]{appendix}

\usepackage{algorithm}
\usepackage{array}
\usepackage{wrapfig}

\usepackage{amsmath}
\usepackage[utf8]{inputenc} 
\usepackage[T1]{fontenc}    
\usepackage{hyperref}       
\usepackage{url}            
\usepackage{booktabs}       
\usepackage{amsfonts}       
\usepackage[most]{tcolorbox}
\usepackage{nicefrac}       
\usepackage{microtype}      
\usepackage{xcolor} 
\usepackage{amssymb}
\usepackage{caption}
\usepackage{xspace}
\usepackage{multirow}
\usepackage{xcolor}
\usepackage{graphicx} 

\usepackage{xspace}
\usepackage{multirow}
\usepackage{seqsplit}

\usepackage{tcolorbox}
\usepackage{mdframed}
\mdfdefinestyle{insightbox}{
  backgroundcolor=green!5!white,
  linecolor=green!40!black,
  linewidth=0.3pt,
  roundcorner=2pt,
  innerleftmargin=4pt,
  innerrightmargin=4pt,
  innertopmargin=3pt,
  innerbottommargin=3pt
}
\mdfdefinestyle{motivationbox}{
  backgroundcolor=blue!5!white,
  linecolor=blue!40!black,
  linewidth=0.3pt,
  roundcorner=2pt,
  innerleftmargin=4pt,
  innerrightmargin=4pt,
  innertopmargin=3pt,
  innerbottommargin=3pt
}
\mdfdefinestyle{codebox}{
  backgroundcolor=black!4!white,
  linecolor=black!40!black,
  linewidth=0.3pt,
  roundcorner=2pt,
  innerleftmargin=4pt,
  innerrightmargin=4pt,
  innertopmargin=3pt,
  innerbottommargin=3pt
}

\usepackage[utf8]{inputenc} 
\usepackage[T1]{fontenc}    
\PassOptionsToPackage{hyphens}{url}
\usepackage{hyperref}       
\usepackage{xurl}           
\usepackage{booktabs}       
\usepackage{amsfonts}       
\usepackage{nicefrac}       
\usepackage{microtype}      
\usepackage{xcolor}         %
\usepackage{xspace}
\usepackage{amssymb}
\usepackage{tabularx}
\usepackage{caption}

\usepackage{amsmath}
\usepackage{algorithm}
\usepackage{algpseudocode}
\usepackage{graphicx}
\usepackage{subcaption}
\usepackage{wrapfig}

\ifdefined\final
\usepackage[disable]{todonotes}
\else
\usepackage[textsize=tiny]{todonotes}
\fi

\usepackage{amsmath,amssymb,bbm,mathtools}
\usepackage{graphicx,subcaption,booktabs,multirow}
\usepackage{xspace,microtype}
\usepackage{tikz}
\usepackage{hyperref}
\usepackage{algorithm}
\usepackage{wrapfig}
\usepackage{algpseudocode}
\usepackage{mdframed}
\mdfdefinestyle{insightbox}{
  backgroundcolor=green!5!white,
  linecolor=green!40!black,
  linewidth=0.3pt,
  roundcorner=2pt,
  innerleftmargin=4pt,
  innerrightmargin=4pt,
  innertopmargin=3pt,
  innerbottommargin=3pt
}

\newcommand{\sys}{\textsc{HexAGenT}\xspace}

\usepackage{natbib}
\usepackage{latexsym}

\usepackage{url}
\usepackage{amssymb}
\usepackage[utf8]{inputenc}
\usepackage{microtype}
\usepackage{booktabs}
\usepackage{pifont} 
\usepackage{multirow}
\usepackage{makecell}
\usepackage{paralist}
\usepackage{xspace}
\usepackage{color}
\usepackage{xcolor}
\usepackage{colortbl}
\usepackage{adjustbox}
\usepackage{hyperref} 
\usepackage[edges]{forest}
\usepackage{tikz} 
\usepackage{caption}
\usepackage{amsfonts}

\hypersetup{
    colorlinks,
    linkcolor={blue!80!black},
    citecolor={blue!80!black},
}
\tikzset{
    root/.style =             {align=center, text width=1cm, rounded corners=3pt, line width=0.3mm, fill=gray!10, draw=gray!80, font=\small},
    demographic/.style =         {align=center, text width=1.8cm, rounded corners=3pt, line width=0.3mm, fill=blue!10, draw=blue!80, font=\footnotesize},
    demographic_work/.style =    {align=center, text width=10cm, rounded corners=3pt, line width=0.3mm, fill=blue!10, draw=blue!0, font=\footnotesize},
    character/.style =         {align=center, text width=1.8cm, rounded corners=3pt, line width=0.3mm, fill=red!10, draw=red!80, font=\footnotesize},
    character_work/.style =    {align=center, text width=10cm, rounded corners=3pt, line width=0.3mm, fill=red!10, draw=red!0, font=\footnotesize},
    personalization/.style =           {align=center, text width=1.8cm, rounded corners=3pt, line width=0.3mm, fill=cyan!10, draw=cyan!80, font=\footnotesize},
    personalization_work/.style =      {align=center, text width=10cm, rounded corners=3pt, line width=0.3mm, fill=cyan!10, draw=cyan!0, font=\footnotesize},
    risk/.style =         {align=center, text width=1.8cm, rounded corners=3pt, line width=0.3mm, fill=orange!10, draw=orange!80, font=\footnotesize},
    risk_work/.style =    {align=center, text width=10cm, rounded corners=3pt, line width=0.3mm, fill=orange!10, draw=orange!0, font=\footnotesize},
}

\usepackage{CJK}

\newtcolorbox{promptbox}[1][]{
  enhanced,
  breakable,
  colback=promptboxlightgray,
  colframe=promptboxblue!30,
  arc=8pt,
  boxrule=0.5pt,
  left=12pt,
  right=12pt,
  top=8pt,
  bottom=8pt,
  fonttitle=\bfseries,
  fontupper=\linespread{1.2}\selectfont,
  title=#1
}

\title{\sys: Efficient Agentic LLM Serving via Workflow- and Heterogeneity-Aware Scheduling}

\author{You Peng$^1$$^*$, Youhe Jiang$^1$$^*$, Wenshuang Li$^1$, Xu Xu$^1$, Ke Zhou$^2$, \\Jiawei Jiang$^3$, Chen Wang$^4$, Binhang Yuan$^1$}

\affiliation{$^1$HKUST, $^2$Webank, $^3$Wuhan University, $^4$Tsinghua University}

\contribution{$^*$Equal contribution}

\abstract{
Agentic LLM applications increasingly execute user requests as multi-step workflows involving planning, tool use, branching, refinement, and synthesis. In such settings, users experience the end-to-end latency of an entire workflow, not the latency of any single LLM call. In this paper, we study how to schedule online agentic workflows across heterogeneous prefill-decode disaggregated LLM serving clusters to efficiently meet workflow-level latency objectives. The problem is challenging because workflow dependencies are revealed incrementally at runtime, calls have heterogeneous prompts, outputs, and KV-cache requirements, and the prefill and decode stages impose different compute, memory, and transfer constraints across heterogeneous GPUs. To solve this problem, we present \sys, a workflow-aware scheduler for a heterogeneous prefill-decode inference service. \sys models each request as an online-revealed DAG, maintains a running estimate of the workflow’s standalone completion horizon, prioritizes ready calls by projected risk of missing that horizon, and jointly selects prefill placement, decode placement, and local queue priority while accounting for KV-cache capacity and cross-stage transfer latency. Across representative agentic workloads and heterogeneous A100/H100/H200 clusters, \sys reduces the SLO scale required for timely workflow completion by an average of $20.1\%$ at $95\%$ attainment and $33.0\%$ at $99\%$ attainment, with maximum reductions of $45.0\%$ and $80.5\%$, respectively.
}

\begin{document}

\maketitle

\section{Introduction}

Large language models are increasingly deployed not as isolated single-turn services, but as components inside agentic workflows: multi-step executions in which an agent plans~\cite{yao2023tree,yao2023react}, invokes tools~\cite{schick2023toolformer,yao2023react}, refines intermediate outputs~\cite{madaan2023selfrefine}, branches into parallel sub-tasks~\cite{yao2023tree,besta2024graph}, and eventually synthesizes a final response. For example, ReAct-style~\cite{yao2023react} agents interleave reasoning with external actions, LATS-style~\cite{zhou2024lats} agents introduce tree-search structure, and function-calling benchmarks such as BFCL~\cite{patil2025bfcl} make clear that tool use and multi-step execution are now central to practical LLM applications. As a result, a single user request often expands into a multi-call workflow whose latency is determined by interactions among many dependent LLM calls rather than by any single call in isolation. This shift changes the fundamental unit of serving. Users do not experience the latency of an individual LLM call in isolation---they experience the end-to-end completion time of the entire workflow. In this paper, we explore \textit{how we should schedule agentic workflows over one state-of-the-art economic LLM serving infrastructure (i.e., prefill-decode disaggregated deployment over heterogeneous compute resources) so that the entire workflow, rather than its individual calls, completes within an optimal end-to-end latency target}.

\vspace{0.25em}
Addressing this problem directly benefits various agentic LLM workflows by improving performance and economic efficiency. \underline{First}, \textit{agentic AI is rapidly moving into production settings}: McKinsey’s 2025 survey~\cite{the_state_of_AI_in_2025} reports that $23\%$ of organizations are already scaling at least one agentic AI system and another $39\%$ are experimenting with them, while the broader economic potential of generative AI is measured in the trillions of dollars annually. \underline{Second}, \textit{heterogeneous deployment is economically attractive}: in modern LLM serving, prefill and decode stress different hardware resources: prefill is compute-intensive, while decode is memory- and KV-cache-intensive. Prefill–decode disaggregation~\cite{zhong2024distserve,patel2024splitwise} therefore creates an opportunity to specialize resources, improve utilization, and reduce the cost of serving. This opportunity is particularly important in heterogeneous deployments, where providers may combine multiple GPU generations, e.g., A100, H100, and H200, to reuse existing accelerator fleets while selectively exploiting newer, more capable hardware. If scheduled well, such heterogeneous deployments can improve the economics of agentic serving by delivering workflow-level SLOs without requiring every request to run on the newest or fastest accelerator.

\vspace{0.25em}
However, agentic workflow serving is difficult because the scheduler must reason simultaneously about online uncertainty, heterogeneous resources, and workflow structure. \underline{First}, the workflow is revealed \textit{online with uncertainty}: imagining LLM call dependence as a direct acyclic graph (DAG), only source calls are visible at arrival, and downstream calls appear only after their parents complete or tools return, so the scheduler must act under partial information. \underline{Second}, workflow structure is coupled with \textit{resource heterogeneity}: calls within the same workflow can have very different prompt lengths, output lengths, and KV-cache demands, while prefill and decode instances may differ substantially in compute speed, memory capacity, and transfer cost. \underline{Third}, local call-level scheduling objectives can conflict with end-to-end \textit{workflow structure}---naive call-level policies optimize the wrong target. A policy such as FCFS~\cite{yu2022orca,kwon2023vllm} or queue-length balancing~\cite{azar1994balanced} can reduce local waiting time for individual calls, yet still delay the critical path of a workflow and harm end-to-end SLO attainment.

\vspace{0.25em}
As a result, existing LLM-serving systems cannot support agentic workflow serving efficiently and economically. Systems such as vLLM~\cite{kwon2023vllm} and SGLang~\cite{zheng2024sglang} provide efficient execution substrates through batching and KV-cache management, but they are primarily designed around individual requests or calls rather than workflow-level SLO objectives. Prefill–decode disaggregated systems such as DistServe~\cite{zhong2024distserve} and Splitwise~\cite{patel2024splitwise} improve phase specialization, but they do not directly solve the problem of scheduling dependency-constrained, online-revealed agentic workflows across heterogeneous Prefill–Decode pools. Program-aware systems such as Parrot~\cite{lin2024parrot}, Hermes~\cite{liu2026hermes}, Autellix~\cite{luo2025autellix}, and Continuum~\cite{li2025continuum} move toward application-level scheduling, but existing approaches do not jointly handle online-revealed workflow DAGs, heterogeneous prefill/decode placement, decode-capacity constraints, stage-coupled planning, and asynchronous runtime overheads. Thus, prior work provides important building blocks, but not a scheduler that is simultaneously workflow-aware, heterogeneity-aware, and tightly coupled to disaggregated serving dynamics.

\vspace{0.25em}
We present \sys, a workflow-aware scheduler for agentic LLM serving on heterogeneous prefill–decode clusters. \sys abstracts each request as an online-revealed DAG and maintains, for every workflow, a running estimate of the standalone completion time implied by the revealed subgraph; this estimate serves as the current planning horizon for the workflow. Ready calls are then prioritized by their projected risk of violating that horizon. To turn this signal into concrete actions, \sys jointly chooses prefill placement, decode placement, and local queue priority, while accounting for decode KV-capacity constraints, cross-stage transfer latency, and asynchronous runtime semantics. Concretely, we make the following contributions:

\begin{itemize}
\item We present \textbf{a representative case study} of agentic workflows to identify the system bottlenecks. We characterize why per-call serving policies are insufficient for agentic workloads and show that workflow-level ordering and heterogeneous placement are both necessary for tight end-to-end SLO attainment.

\item We design \textbf{a workflow- \& heterogeneity-aware scheduling algorithm}, where we formulate agentic serving as an online workflow scheduling problem over heterogeneous resources and design \sys, which combines online workflow horizons, projected-risk prioritization, joint prefill–decode placement, decode-capacity-aware scheduling, and asynchronous plan application.

\item We show the \textbf{performance boost} of \sys through an empirical study across representative agentic workloads and heterogeneous clusters. Across heterogeneous A100/H100/H200 settings, \sys lowers the SLO scale required to complete workflows on time relative to the strongest baselines, with especially large gains on mixed and tail-heavy workloads. The current manuscript’s headline result reports average reductions of $20.1\%$ for $95\%$ workflow attainment and $33.0\%$ for $99\%$ workflow attainment, with maximum reductions of $45.0\%$ and $80.5\%$.

\end{itemize}

\section{Preliminaries and Related Work}
\label{sec:preliminaries}

\noindent\textbf{LLM serving systems.}
\label{subsec:prelim-llm-serving}
Modern LLM serving systems are primarily designed around \emph{individual inference requests} as the basic scheduling unit. Their main optimizations focus on memory-efficient KV-cache management, continuous batching, and high-throughput execution on shared accelerators. Representative systems such as ORCA~\cite{yu2022orca}, vLLM~\cite{kwon2023vllm}, SGLang~\cite{zheng2024sglang}, Sarathi-Serve~\cite{agrawal2024sarathiserve}, Llumnix~\cite{sun2024llumnix}, and ServerlessLLM~\cite{fu2024serverlessllm} have demonstrated that careful runtime design can substantially improve throughput and tail latency for conventional single-request or multi-tenant inference workloads. Nevertheless, these systems largely assume that requests are independent, or at least that queueing decisions can be made at the granularity of individual calls.
This request-centric design is well aligned with chatbot-style workloads, but it becomes insufficient for agentic applications. In agentic execution, multiple LLM calls belong to the same end-to-end user task, and their value depends on workflow context rather than on standalone completion time. As a result, scheduling policies that are effective for independent LLM calls may still perform poorly once user requests expand into dependency-constrained multi-call workflows.

\nocite{peng2025hexgen,jiang2023hexgen,jiang2025hexgen,jiang2025demystifying}

\vspace{0.5em}
\noindent\textbf{Prefill--decode disaggregation.}
\label{subsec:prelim-pd}
Prefill--decode (P-D) disaggregation has emerged as an important architecture for large-scale LLM serving~\cite{zhong2024distserve,jin2024pdserve,agrawal2024sarathiserve}. The motivation is that the two phases of autoregressive inference stress different resources: prefill is compute-intensive and benefits from high-throughput execution, whereas decode is memory- and KV-cache-intensive and is often bottlenecked by capacity and admission decisions. Separating these two phases onto dedicated instance pools can therefore improve hardware specialization and resource utilization.
At the same time, P-D disaggregation makes scheduling more difficult. A single LLM call is no longer mapped to one monolithic worker; instead, the system must decide which prefill instance should process the prompt, which decode instance should continue generation, and when the call should enter each local queue. The serving path also includes an explicit transfer stage between prefill and decode. Even when transfer is not itself scheduled as a separate shared bottleneck, it still contributes non-negligible latency and affects when decode becomes ready. Thus, disaggregation transforms serving into a multi-stage scheduling problem with stage coupling.

\nocite{miao2022galvatron,jiang2022osdp,wang2024improving,he2026efficient,yan2025areal,yan2025fsa,he2026efficient2}

\vspace{0.5em}
\noindent\textbf{Program-aware and agentic serving.}
\label{subsec:prelim-program-aware}
Recent work has begun to recognize that many LLM applications are better viewed as \emph{programs} rather than as bags of independent requests. Systems and schedulers such as Parrot~\cite{lin2024parrot}, Hermes~\cite{liu2026hermes}, Autellix~\cite{luo2025autellix}, and Continuum~\cite{li2025continuum} explicitly elevate the scheduling abstraction from an individual request to a larger program, application, or multi-turn agent execution. Program-aware schedulers, such as program-level FCFS and the attained-service-based PLAS/ATLAS family, treat the whole program or workflow as the primary object of fairness and progress accounting. This line of work is especially relevant for agentic workloads, where a user request may execute as a chain, tree, or DAG of dependent LLM calls.
However, existing program-aware schedulers do not fully match the setting studied in this paper. Our target system combines three properties simultaneously: \textbf{(i)} workflows are revealed online rather than fully known in advance, \textbf{(ii)} serving is P-D disaggregated with heterogeneous prefill and decode instances, and \textbf{(iii)} the scheduler must operate under realistic runtime overhead constraints and therefore cannot block the serving path. These differences motivate a scheduler that is both workflow-aware and tightly coupled to the dynamics of P-D disaggregated execution.

\nocite{jiang2026autopoiesis}

\vspace{0.5em}
\noindent\textbf{Call-Level scheduling for LLM inference.}
\label{subsec:related-call}
Another related line of work studies call-level scheduling policies for LLM serving, including preemptive queueing, shortest-job-first variants, deadline- or slack-aware policies, attained-service policies, and output-length-prediction-based scheduling. FastServe~\cite{wu2023fastserve} applies iteration-level preemptive scheduling and multi-level feedback queues to reduce head-of-line blocking. QLM~\cite{patke2024qlm} studies queue management for SLO-oriented LLM serving. SSJF~\cite{qiu2024ssjf} uses proxy-model-based sequence length prediction for speculative shortest-job-first scheduling, while LTR-based scheduling~\cite{fu2024ltr} and embedding-based scheduling~\cite{shahout2024dontstop} use learned signals to improve prioritization. BlendServe~\cite{zhao2024blendserve} further explores resource-aware batching for autoregressive inference. These methods improve per-request latency under contention and provide useful building blocks for practical serving systems.
However, these policies remain blind to workflow structure. When multiple calls belong to the same agentic request, optimizing them independently can still produce poor end-to-end behavior. A scheduler may reduce the latency of isolated calls while delaying a workflow's critical path or starving large but important workflows.

\nocite{jiang2025thunderserve,tong2025parallax,jiang2026boute}

\vspace{0.5em}
\noindent\textbf{Classical scheduling and DAG scheduling.}
\label{subsec:related-classical}
Our work is also related to classical disciplines such as shortest-job-first, shortest-remaining-processing-time, least-slack-time-first, and least-attained-service scheduling~\cite{schrage1968proof,davis1993slack}. These policies provide useful principles for prioritization, fairness, and deadline awareness, and they inspire several of our baselines. We are also informed by the broader literature on workflow and DAG-style execution in agentic systems, including ReAct-style tool-using agents~\cite{yao2023react}, tree-structured reasoning agents such as LATS~\cite{zhou2024lats}, and agentic tool-use benchmarks such as BFCL~\cite{patil2025bfcl}.
Agentic LLM serving combines several properties that are uncommon in classical settings: workflow structure is revealed online, call durations are only estimated, decode memory is a first-class capacity constraint, and resource choice is phase-disaggregated. These differences motivate a scheduler specialized for agentic workflows rather than a direct reuse of generic job scheduling methods.

\vspace{0.5em}
\noindent\textbf{Heterogeneous LLM serving.}
\label{subsec:heterogeneous-llm-serving}
Heterogeneous LLM serving has become increasingly important as production clusters often combine different GPU generations, memory capacities, network bandwidths, cloud regions, and pricing models. Several recent systems~\cite{jiang2025hexgen,mao2025skyserve,jiang2025thunderserve} exploit such heterogeneity to improve throughput, latency, availability, or cost efficiency. For instance, HexGen~\cite{jiang2023hexgen} serves LLMs over heterogeneous GPUs and networks through asymmetric tensor/pipeline partitioning and constrained scheduling. Splitwise~\cite{patel2024splitwise} exploits the distinct resource characteristics of prompt processing and token generation by splitting them across phase-specialized machines, including heterogeneous cluster designs.
These systems demonstrate the importance of heterogeneity in LLM serving, but they primarily optimize placement or request scheduling at the serving-system level. In contrast, our work treats heterogeneity as part of the workflow scheduling problem: the scheduler reasons not only about which call is urgent, but also about which prefill/decode placement best reduces the workflow's projected scaled-SLO risk under heterogeneous service times.

\nocite{jiang2026oserve,jiang2025cascadia,zhang2025efficient}


\section{Workload Characterization}
\label{sec:characterization}

This characterization section motivates why serving agentic workflows on heterogeneous P-D clusters requires more than per-call level scheduling. We conduct a small ablation study to answer two questions: \textbf{(i)} whether end-to-end workflow awareness is necessary for agentic serving, and \textbf{(ii)} whether workflow awareness alone is sufficient without any heterogeneous workload/GPU-aware placement. This section is meant to isolate the design motivations behind \sys, and we will comprehensively compare all workflow-level schedulers in Section~\ref{sec:evaluation}.

\vspace{0.5em}
\noindent\textbf{Experiment setting.}
We evaluate on two models: \textsc{Llama3.1-70B} and \textsc{Qwen3-235B-A22B}. The experiments use a heterogeneous P-D cluster with $8$ prefill instances and $8$ decode instances. Each pool is composed of $2$ A100, $3$ H100, and $3$ H200 model instances. The workloads include ShareGPT-style conversational chains, BFCL-v3 function-calling workflows, and LATS-style tree/search workflows.

\vspace{0.5em}
\noindent\textbf{Metrics.} We use the scaled-SLO metric for evaluation. For each workflow $w$, we first estimate its standalone execution horizon $H_w$ on the same serving cluster. Given a scale factor $\alpha$, the workflow is counted as satisfying its SLO if its completion time $C_w$ satisfies $C_w \le \alpha H_w$. We report Req95 and Req99, which denote the minimum scale factor $\alpha$ required for $95\%$ and $99\%$ workflow SLO attainment, respectively. Lower Req95/Req99 indicates that the same serving system can meet workflow SLOs under tighter latency targets.

\vspace{0.5em}
\noindent\textbf{Baseline system configuration.}
We begin with a representative LLM serving system as the baseline: SGLang with per-call FCFS as default scheduling policy. This baseline treats each revealed LLM call as an independent request. It does not account for the fact that multiple calls belong to the same end-to-end workflow, and it does not use heterogeneous per-instance latency estimates when deciding where a call should run. As shown in Table~\ref{tab:characterization-ablation}, this baseline often requires large SLO scales to satisfy tail workflows. For example, on Qwen-BFCL, it requires a Req95 of $21.11$ and a Req99 of $26.89$; on Llama-BFCL, it requires a Req95 of $13.81$ and a Req99 of $17.23$.

\vspace{0.5em}
\noindent\textbf{\underline{Approach 1:} Adding workflow-level awareness.}
The first improvement is to replace per-call FCFS with workflow-level FCFS. Instead of ordering all calls purely by reveal time, this policy preserves the ordering of the workflow. This simple change already improves end-to-end behavior because it prevents calls from the same workflow from being arbitrarily interleaved behind calls from unrelated workflows. Across the six cases in Table~\ref{tab:characterization-ablation}, Workflow-FCFS reduces Req95 by $31.4\%$ on average and Req99 by $23.3\%$ on average compared with per-call FCFS. The improvement is especially large on BFCL: for Qwen-BFCL, Req95 drops from $21.11$ to $9.64$, and Req99 drops from $26.89$ to $11.67$.

\begin{table}[t]
  \centering
  \caption{Characterization ablation on heterogeneous P-D deployments. Each cell reports Req95/Req99. The table progressively adds workflow-level ordering and then \sys's heterogeneous workload/GPU-aware scheduling. Lower is better.}
  \label{tab:characterization-ablation}
  \vspace{-0.5em}
  \resizebox{0.6\columnwidth}{!}{%
  \begin{tabular}{l|l|c|c|c}
  \hline
  \textbf{Model} & \textbf{Trace} & \textbf{Per-call FCFS} & \textbf{Workflow-FCFS} & \textbf{\sys} \\
  \hline
  Llama & ShareGPT & 5.85 / 7.43 & 4.50 / 6.22 & 2.50 / 2.60 \\
  \hline
  Llama & BFCL-v3  & 13.81 / 17.23 & 7.23 / 9.80 & 6.21 / 6.34 \\
  \hline
  Llama & LATS     & 3.21 / 3.29 & 2.37 / 3.03 & 2.19 / 2.24 \\
  \hline
  Qwen  & BFCL-v3  & 21.11 / 26.89 & 9.64 / 11.67 & 8.39 / 8.57 \\
  \hline
  Qwen  & LATS     & 4.55 / 4.78 & 3.21 / 4.27 & 2.71 / 2.74 \\
  \hline
  Qwen  & Mixed    & 11.15 / 15.84 & 10.30 / 15.01 & 3.48 / 3.94 \\
  \hline
  \end{tabular}%
  }
  \vspace{-0.5em}
\end{table}

\vspace{0.5em}
\begin{mdframed}[style=insightbox]
\small
\noindent\textbf{\underline{Insight 1.}} Agentic workflow serving must optimize for end-to-end workflow progress, not only per-call queue order. Even a simple workflow-level FCFS policy substantially improves tail SLO scales over per-call FCFS because it preserves workflow-level execution continuity.
\end{mdframed}
\vspace{0.5em}

\vspace{0.5em}
\noindent\textbf{\underline{Approach 2:} Adding heterogeneity awareness.}
Adding workflow-level ordering is not sufficient. Agentic workflows contain calls with different input lengths and output lengths. Some calls are more prefill-heavy, while others create larger decode/KV pressure (i.e., workload heterogeneity~\cite{zhong2024distserve}). In a heterogeneous P-D cluster, different model instances also incur different prefill/decode service latencies (i.e., hardware heterogeneity~\cite{jiang2025hexgen}). Therefore, after deciding which workflow is urgent, the scheduler must also choose an appropriate prefill/decode pair for each revealed call.

\vspace{0.25em}
\sys adds this second dimension. It uses the current workflow SLO target to prioritize ready calls and then evaluates candidate prefill/decode placements using estimated per-instance service latencies and current queue states. Compared with Workflow-FCFS, \sys further reduces Req95 by $26.9\%$ on average and Req99 by $42.6\%$ on average across the six characterization cases. The largest improvement appears on Qwen-Mixed, where Req95 drops from $10.30$ to $3.48$ and Req99 drops from $15.01$ to $3.94$. This shows that workflow-level ordering alone cannot fully exploit heterogeneous resources; the scheduler must also reason about which model instance can serve each call most effectively.

\vspace{0.5em}
\begin{mdframed}[style=insightbox]
\small
\noindent\textbf{\underline{Insight 2.}} End-to-end workflow awareness must be combined with heterogeneous placement. A scheduler should not only know which workflow is urgent, but also account for the workload pattern of each LLM call and the expected service time on each prefill/decode instance.
\end{mdframed}
\vspace{0.5em}

The two insights together motivate the design of \sys. It treats agentic LLM serving as a joint workflow scheduling and heterogeneous placement problem. It first reasons about each workflow's current progress toward its end-to-end SLO target, and then assigns each ready call to the prefill/decode instances that best reduce the workflow's projected scaled-SLO pressure. This design improves the cost efficiency of agentic workflow serving: under the same heterogeneous P-D deployment, the system can reach the same workflow-level SLO attainment at a much smaller SLO scale. We next present the system design in Section~\ref{sec:overview} and the scheduling algorithm in Section~\ref{sec:scheduling}.

\section{System Overview}
\label{sec:overview}

This section presents the architecture of \sys and clarifies the scheduler's role during serving. The key idea is to treat an agentic request as an online-revealed workflow while serving it in a P-D disaggregated runtime.

\subsection{Agentic Workflow Abstraction as an Online DAG}
\label{subsec:overview-dag}
We first define how \sys represents an agentic workflow as a DAG, and then explain why this DAG should be treated as an online-revealed structure.

\begin{wrapfigure}{r}{0.45\linewidth}
\vspace{-0.5em}
\centering
\includegraphics[width=\linewidth]{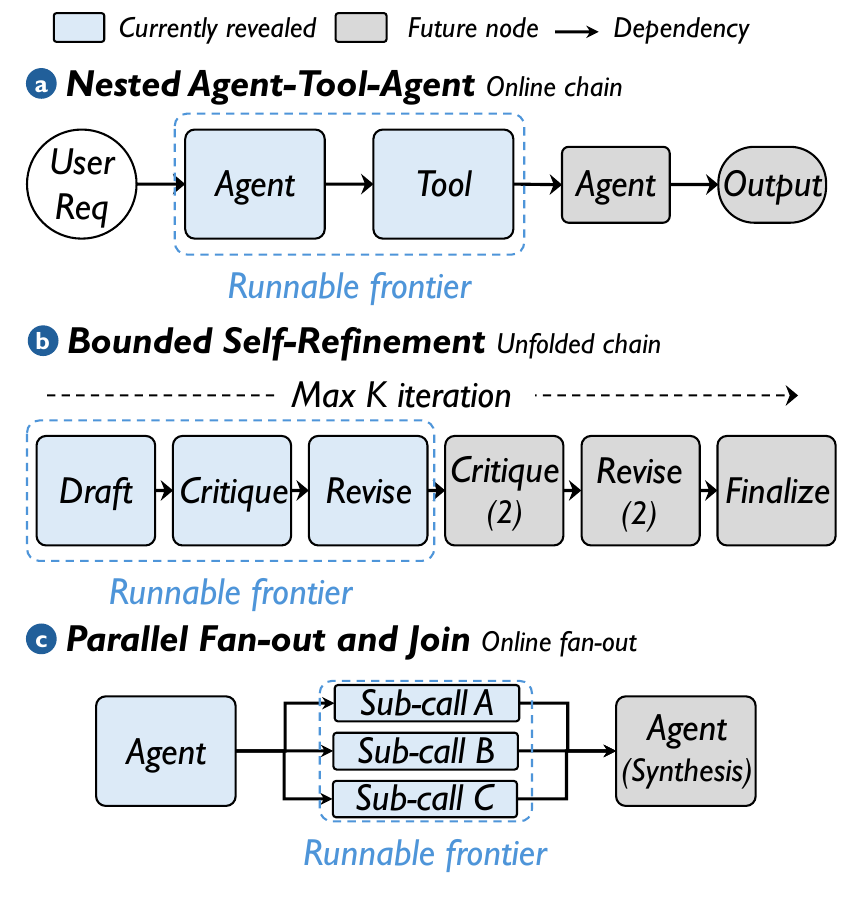}
\caption{Example of an agentic LLM application workflow. Only source calls are known at arrival. As parent calls and tool calls complete, new LLM calls are revealed, and the workflow DAG grows online.}
\label{fig:workflow_reveal}
\vspace{-1.0em}
\end{wrapfigure}

\vspace{0.5em}
\noindent\textbf{Workflow DAG abstraction.} \sys abstracts each agentic workflow as a DAG $G=(V,E)$ whose nodes are LLM calls and whose edges encode execution dependencies. This abstraction captures the common control-flow patterns of agentic applications. A nested agent-tool-agent interaction appears as a short chain; a bounded self-refinement process can be unfolded into a finite chain or subtree; and parallel sub-calls appear as sibling branches that become runnable simultaneously after a common parent finishes.

\vspace{0.5em}
\noindent\textbf{Online reveal and runnable frontier.}
The key property of this abstraction is that the DAG is \emph{revealed online}. At workflow arrival, only source nodes are visible to the scheduler. A child node becomes schedulable only after all of its parents are complete. Therefore, the scheduler never plans the full workflow graph upfront. Instead, the system maintains a runnable frontier, and it repeatedly makes decisions over the currently revealed frontier of runnable calls. This online-reveal semantics matches how real agentic applications execute and is also the abstraction used by our scheduler. Figure \ref{fig:workflow_reveal} presents examples of this revealing process.



\subsection{Serving Architecture and Scheduler Placement}
\label{subsec:overview-architecture}
We first describe the major system components, then explain the internal modules of the scheduler, and finally specify the heterogeneous resources that the scheduler models when making placement and queuing decisions.

\begin{figure*}
    \centering
    \includegraphics[width=0.9\linewidth]{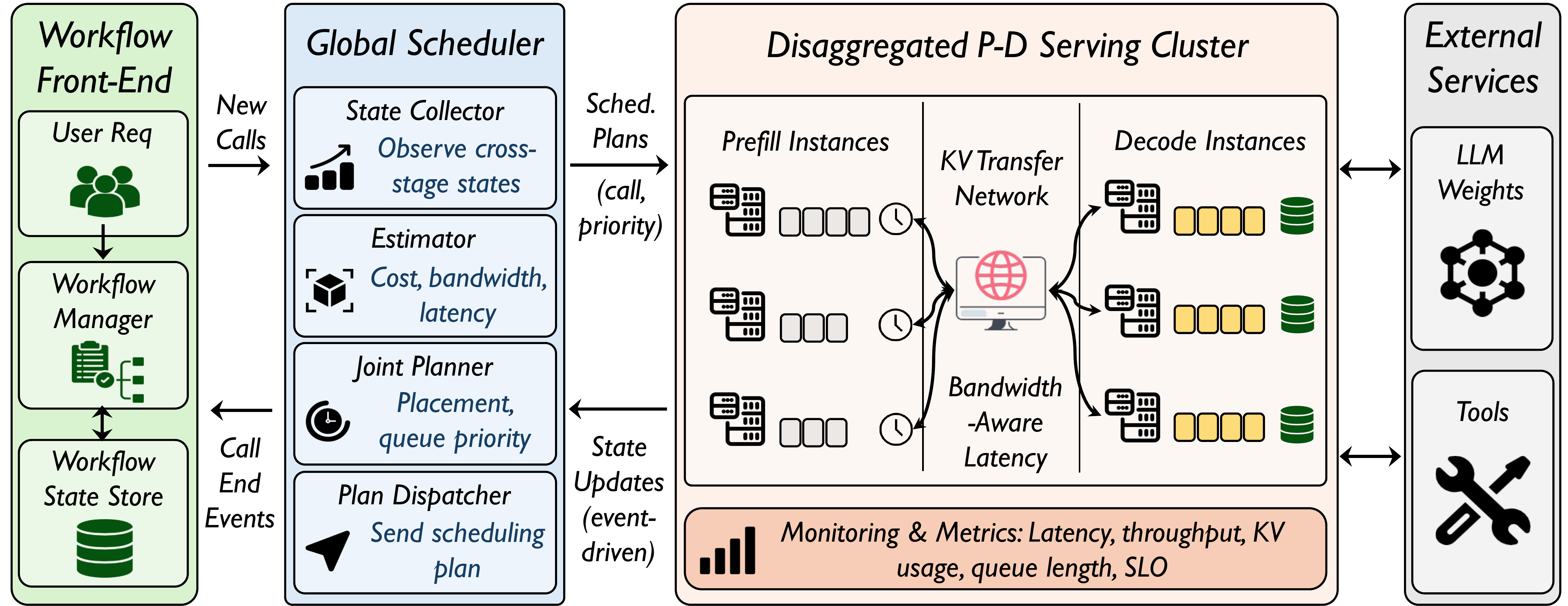}
    \caption{System architecture and scheduler placement of \sys. The workflow front-end releases ready calls from online agent workflows, while the global scheduler collects cross-stage state, estimates prefill, KV-transfer, and decode latencies, and jointly decides instance placement and queue priority. The P-D disaggregated cluster executes calls across prefill and decode stages, with runtime metrics fed back for event-driven re-scheduling. \sys frames agent serving as a joint placement-and-queueing problem over heterogeneous GPU resources to support PD disaggregated deployment.}
    \label{fig:system_archi}
\end{figure*}

\vspace{0.5em}
\noindent\textbf{System components.} Figure \ref{fig:system_archi} presents the architecture of our system. \sys is organized around four logical components: a \textit{workflow front-end}, a \textit{global scheduler}, a \textit{P-D serving cluster}, and \textit{external services}. The \textit{workflow front-end} receives agentic requests, tracks workflow state, and maintains the online DAG reveal process. The \textit{global scheduler} observes the live system state and determines how waiting calls should be assigned and ordered. The \textit{P-D serving cluster} contains a prefill pool and a decode pool built from heterogeneous hardware instances. Finally, \textit{external services} include LLMs and tools that can be called in the workflow.

\vspace{0.5em}
\noindent\textbf{Scheduler internals.} The scheduler is not merely a routing policy. It acts as a live assignment and queue-control policy supported by four internal modules. The \textit{State Collector} gathers cross-stage runtime snapshots, including prefill queues, decode queues, running calls, decode KV usage, bandwidth-related transfer state, and workflow progress. The \textit{Estimator} converts these snapshots and per-call input and output length into runtime estimates using a roofline-style model~\cite{williams2009roofline,cao2025moe_lightning}, including prefill time, decode time, KV-transfer latency, and decode memory demand. The \textit{Joint Planner} then computes the scheduling plan by jointly deciding (i) the prefill/decode pair that should serve each waiting call and (ii) the local queue priority that the call should receive on the chosen instances. Finally, the \textit{Plan Dispatcher} executes the scheduling decisions by sending placement and priority updates to the corresponding prefill and decode workers. This distinction is important because queue order, not just the choice of model instance, strongly affects workflow tail latency in disaggregated serving.

\vspace{0.5em}
\noindent\textbf{Heterogeneous resource modelling.} \sys explicitly models the heterogeneous resources in a P-D disaggregated serving cluster. Each prefill instance behaves as a single-server execution engine with a local waiting queue, while each decode instance performs batched decoding subject to a KV-cache capacity constraint. The serving path also includes an explicit KV-transfer stage between prefill and decode. We compute the transfer latency from the KV size and the measured effective bandwidth between the corresponding GPU classes; in heterogeneous clusters, cross-hardware transfers therefore naturally incur higher latency when the available bandwidth is lower. The scheduler reasons over prefill queues, decode queues, decode KV capacity, and transfer-induced decode-ready times when making placement and ordering decisions.

\subsection{End-to-End Request Lifecycle}
\label{subsec:overview-lifecycle}
We first introduce how LLM calls are revealed and scheduled, then specify when the scheduling is triggered and how the scheduled plan is applied.

\vspace{0.5em}
\noindent\textbf{Workflow arrival reveals the initial runnable calls.} An end-to-end agentic request begins when the workflow front-end receives a new user request and instantiates its workflow state. Only source LLM calls are revealed upon workflow arrival. These LLM calls are inserted into the global waiting set for prefill scheduling. Before invoking the scheduler, the front-end gathers a live snapshot of the system state, including current prefill and decode queues, running calls, decode capacities, and workflow-level progress metadata.

\vspace{0.5em}
\noindent\textbf{Scheduler assigns instance and queue priorities.} The scheduler then produces a plan for the currently waiting calls. For each call, the plan includes a chosen prefill instance, a chosen decode instance, and stage-local priorities. The call is dispatched to the selected prefill instance, enters a short bootstrap process with the chosen decode instance, and then joins that prefill instance's local waiting queue. Once prefill starts, the call is no longer movable across prefill instances. After the prefill finishes, the system performs KV transfer to the selected decode instance. When the transfer completes, the call becomes decode-ready and enters the decode-side waiting structures of the chosen decode instance. Decode then begins subject to memory capacity and local queue order. When decode finishes, the call is marked complete, the workflow state is updated, and any newly unblocked child calls are revealed.

\vspace{0.5em}
\noindent\textbf{Runtime events drive subsequent schedulings.} This lifecycle is \emph{event-driven}. The main events are workflow arrival, prefill completion, transfer completion, and decode completion. Each of these events changes the runtime state. Workflow arrival adds newly runnable source calls. Prefill completion releases a prefill slot and initiates transfer. Transfer completion makes a call eligible for decode admission. Decode completion releases decode capacity and may reveal additional child calls. In the current implementation, scheduler invocations are triggered when new work enters a serving stage---workflow arrival, newly revealed prefill work, and transfer completion into decode.

\vspace{0.5em}
\noindent\textbf{Asynchronous planning avoids blocking.} Asynchronous scheduling is integrated into this lifecycle. While a new scheduling solve is in flight, serving does not stop. Newly arrived waiting calls that still have no assignment may temporarily follow a safe fallback policy. If the asynchronous plan returns before such a call starts service, the waiting call can still be updated to the new plan. If service has already started, the runtime state becomes authoritative, and the late plan is ignored for that serving stage. In short, \sys plans continuously in the background while the foreground serving path keeps making progress.

\section{Scheduling Framework}
\label{sec:scheduling}

This section formalizes the scheduling objective of \sys and then describes the online scheduling algorithm used in our framework. We first state the end-to-end scaled-SLO objective and then explain how \sys makes stage-level decisions under online DAG revelation and P-D disaggregated serving.

\subsection{Problem Formulation}
\label{subsec:formulation}

\vspace{0.5em}
\noindent\textbf{Define online workflow horizon.} We consider a stream of agentic workflows $\mathcal{W}=\{w_1,w_2,\dots\}$ arriving over time. Each workflow $w$ is an online-revealed DAG. Let $G_w(t)$ denote the subgraph of workflow $w$ that has been revealed by time $t$. \sys assigns each workflow a workflow-specific horizon $H_w(t)$, which serves as the current end-to-end serving target during online scheduling. Concretely, $H_w(t)$ is the standalone completion time of $G_w(t)$ on the evaluated P-D cluster: the time needed to execute the revealed workflow subgraph if it were served in isolation using the same serving cluster. When the workflow is fully revealed, the final horizon $H_w$ becomes the standalone completion time of the entire workflow.

\vspace{0.5em}
\noindent\textbf{Optimizing for scaled-SLO objective.} Let $C_w^\pi$ denote the end-to-end completion time of workflow $w$ under scheduling policy $\pi$. Our goal is to maximize end-to-end SLO attainment across workflows. Equivalently, for a target attainment level $\tau \in \{0.95, 0.99\}$, we formulate the objective as finding the smallest scale factor $\alpha$ such that at least a $\tau$ fraction of workflows complete within $\alpha H_w$:
\begin{equation}
\min_{\pi}\; \alpha
\quad
\text{s.t.}
\quad
\frac{1}{|\mathcal{W}|}
\sum_{w\in\mathcal{W}}
\mathbbm{1}\!\left[C_w^\pi \le \alpha H_w\right]
\ge \tau.
\label{eq:e2e_scaled_slo}
\end{equation}
This formulation directly captures the goal of completing more workflows within tighter scaled SLOs, rather than merely reducing average call latency.

\vspace{0.5em}
\noindent\textbf{Refine the horizon progressively.} $H_w$ is not known when a workflow first arrives. Since the workflow DAG is revealed online, the scheduler only observes $G_w(t)$ at time $t$. \sys therefore maintains an online planning horizon $H_w(t)$ and recomputes it whenever new calls are revealed. If a newly revealed call changes the isolated makespan of the revealed subgraph, then the workflow's SLO target grows accordingly. When true service times become available, the horizon can also be corrected by replacing estimates with observed serving latencies. Thus, $H_w(t)$ is a progressively refined estimate of the final workflow-specific target.

\vspace{0.5em}
\noindent\textbf{Introduce the projected ratio as an surrogate.} Since $H_w$ need to be progressively refined, the scheduler does not optimize Equation \eqref{eq:e2e_scaled_slo} directly at every event. Instead, at each scheduling invocation it estimates how scheduling a ready call would affect the workflow's SLO violation risk. For a ready call $c$ from workflow $w$ at time $t$, let $\Delta_s(c,t)$ denote the projected elapsed time from now until the call finishes at stage $s\in\{\mathrm{Prefill},\mathrm{Decode}\}$. For prefill scheduling, this projection includes waiting time, prefill time, and KV-transfer latency. For decode scheduling, it includes waiting for decode capacity and decode execution on a candidate decode instance. \sys defines the projected ratio as
\begin{equation}
R_s(c,t)
=
\frac{(t-a_w)+\Delta_s(c,t)}{H_w(t)},
\label{eq:projected_ratio}
\end{equation}
where $a_w$ is the workflow arrival time, and the term $(t-a_w)$ represents the workflow's elapsed time. A larger $R_s(c,t)$ indicates that serving $c$ is more urgent because the workflow is projected to be closer to, or further beyond, its current serving target.

\subsection{Global Scheduling Loop}
\label{subsec:solver}

The scheduling loop of \sys is event-driven. The runtime state changes on workflow arrival, prefill completion, transfer completion, and decode completion. By design, the scheduler is triggered on the following events: when a workflow arrives and reveals source calls, when decode completion reveals new child calls for prefill, and when transfer completion makes a call ready for decode. Other events still update the runtime state, but they do not necessarily invoke a new scheduling.

\vspace{0.5em}
\noindent\textbf{Workflow-aware scheduling.} The scheduling is stage-aware and workflow-aware. A prefill invocation considers all calls currently waiting for prefill; a decode invocation considers all calls currently waiting for decode. In both cases, \sys observes the workflow horizon $H_w(t)$, the workflow's current completion progress, and the live state of the corresponding stage. Scheduling is asynchronous: at most one plan is in flight, serving continues while a plan is being computed, and late plans are applied only to calls that are still waiting. Once a call starts prefill or decode, its prefill and decode placement are fixed, and the call is no longer movable by the scheduler. This workflow-aware design reflects the first motivation of \sys: scheduling decisions should be made with workflow-level awareness, because the serving objective is end-to-end SLO attainment rather than the latency of isolated LLM calls.


\vspace{0.5em}
\noindent\textbf{Event-driven scheduling.} Algorithm~\ref{alg:hexgent-scheduling} summarizes the event-driven scheduling loop in \sys. At a high level, each scheduling invocation ranks waiting calls by projected scaled-SLO risk, greedily selects the most urgent call, updates a simulated resource state, and finally applies the resulting plan asynchronously to calls that are still waiting. The detailed scheduling logic is described in \S\ref{subsec:prefill} and \S\ref{subsec:decode}.

\begin{algorithm}[t]
\small
\caption{Event-driven scheduling in \sys}
\label{alg:hexgent-scheduling}
\begin{algorithmic}[1]
\Require event $e$, time $t$, runtime state $\mathcal{S}(t)$
\State Update queues, running calls, KV state using $e$
\State Update affected workflow horizons $H_w(t)$
\State $\mathcal{T}\leftarrow$ triggered stages from $e$
\Comment{$\mathcal{T}\subseteq\{\mathrm{P},\mathrm{D}\}$}

\ForAll{stage $s\in\mathcal{T}$}
    \If{planner for $s$ is in flight}
        \State Apply fallback plan if needed
        \State \textbf{continue}
    \EndIf

    \State $Q_s\leftarrow$ waiting calls at stage $s$
    \State $\widehat{\mathcal{S}}\leftarrow\mathcal{S}(t)$
    \State $\Pi_s\leftarrow\emptyset$

    \While{$Q_s\neq\emptyset$}
        \ForAll{$c\in Q_s$}
            \State Project best feasible placement for $c$
            \State Obtain projected elapsed time $\Delta_s(c,t)$
            \State $R_s(c,t)\leftarrow ((t-a_{w(c)})+\Delta_s(c,t))/H_{w(c)}(t)$
        \EndFor

        \State $c^\star\leftarrow\arg\max_{c\in Q_s}R_s(c,t)$
        \State Add placement and rank of $c^\star$ to $\Pi_s$
        \State Update simulated state $\widehat{\mathcal{S}}$
        \State $Q_s\leftarrow Q_s\setminus\{c^\star\}$
    \EndWhile

    \State Asynchronously apply $\Pi_s$ to still-waiting calls
\EndFor
\end{algorithmic}
\end{algorithm}


\subsection{Prefill Scheduler}
\label{subsec:prefill}

\vspace{0.5em}
\noindent\textbf{Projected-risk prefill ranking.} The prefill scheduler ranks waiting calls by projected scaled-SLO risk. For each call, it enumerates candidate prefill instances and candidate decode instances, estimates the prefill finish time, adds transfer latency, and projects the earliest decode finish using current decode availability hints. This gives $R_{\mathrm{P}}(c,t)$ in Equation \eqref{eq:projected_ratio}. Calls with larger projected ratios are treated as more urgent. This enumeration reflects the second motivation of \sys: serving must be heterogeneity-aware, since the same LLM call can have different prefill and decode latencies on different model instances.

\vspace{0.5em}
\noindent\textbf{Adaptive greedy prefill assignment.} For small queues, \sys uses a greedy procedure. It repeatedly selects the most urgent pending call, assigns it to the prefill--decode pair with the earliest projected decode finish, updates the simulated prefill and decode availability, and then recomputes priorities for the remaining calls. This recomputation is important because the best choice for the next call depends on assignments already made in the same planning invocation. For larger queues, \sys uses a one-pass ordering by the same risk score to keep overhead bounded.

\vspace{0.5em}
\noindent\textbf{Joint prefill--decode planning.} The prefill scheduler performs joint prefill--decode planning. Its decision for a call contains both the assigned prefill instance and a planned decode instance. The latter is used to establish the downstream decode relation before prefill completes, so that the runtime knows where the KV state should be transferred. In heterogeneous settings, this joint decision also accounts for the smaller bandwidth across different hardware types, which will discourage prefill-decode pairs with expensive KV movement unless their service-time benefit justifies it.


\subsection{Decode Scheduler}
\label{subsec:decode}

\vspace{0.5em}
\noindent\textbf{Decode scheduling with instance locks.} The decode scheduler operates on calls that have completed transfer and are waiting for decode admission. It uses the same projected-ratio score for ranking, while its feasibility model additionally respects decode KV capacity. A call may be \emph{locked} to a planned decode instance selected during prefill scheduling; locked calls can be reordered within that instance but cannot migrate across decode instances. On the other hand, free calls can be placed on any feasible decode instance.

\vspace{0.5em}
\noindent\textbf{Decode KV-capacity feasibility check.} Feasibility is determined by decode memory demand. Let $L_{\mathrm{in}}(c)$ denote the input prompt length of call $c$ in tokens, and let $\widehat{L}_{\mathrm{out}}(c)$ denote its predicted output length. We use proxy model-based sequence length prediction~\cite{qiu2024proxy} to estimate $\widehat{L}_{\mathrm{out}}(c)$ from the model input prompt and model identity. The decode demand of a call is approximated as
\begin{equation}
m(c)=L_{\mathrm{in}}(c)+\widehat{L}_{\mathrm{out}}(c),
\label{eq:decode-demand}
\end{equation}
and a call is feasible on decode instance $d$ only if
\begin{equation}
m(c) \le \mathrm{Cap}(d),
\label{eq:decode-feasible}
\end{equation}
where $\mathrm{Cap}(d)$ is the decode KV capacity of instance $d$.

\vspace{0.5em}
\noindent\textbf{Adaptive greedy decode assignment.} For small decode queues, \sys again uses recomputing greedy scheduling. It repeatedly selects the most urgent call, finds the earliest capacity-feasible start time on each legal decode instance, assigns the call to the instance with the earliest projected finish, and updates the simulated decode intervals. For larger or special-case queues, it falls back to the lightweight ordering described above. Together, the prefill and decode schedulers realize the core design of \sys: workflow-aware projected-risk priority, heterogeneous prefill--decode matching, decode-capacity-aware placement, and asynchronous online refinement.

\section{Implementation}
\label{sec:implementation}

We implement \sys on top of SGLang v0.5.9~\cite{zheng2024sglang}, using SGLang's P-D disaggregated serving feature as the substrate. The implementation keeps the scheduling policy outside GPU kernels and the hot decode loop, and instead integrates \sys into the gateway and worker path.

\vspace{0.25em}
The implementation mainly uses Python, with JSON configuration files for deployment metadata. The main changes include router-side workflow metadata parsing, stage-state construction, asynchronous schedule plan application, revision checks for safe queue mutation, bootstrap metadata injection, and completion feedback accounting. Worker-side changes expose P-D snapshots, accept priority/reassignment updates for still-waiting requests, and report completion telemetry back to the router.

\vspace{0.25em}
We also implement a standalone event-driven simulator as a resource estimator that supports scheduler decisions. The simulator is implemented in Python and contains about 4.6K lines across the runtime, scheduler, and runners. It models the full lifecycle of each LLM call, including \texttt{waiting prefill}, \texttt{prefill}, \texttt{KV transfer}, \texttt{waiting decode}, \texttt{decode}, and \texttt{completion}. The simulator maintains prefill instance availability, decode token-capacity intervals, workflow DAG dependencies, asynchronous scheduling events, and workflow-level SLO state. For each scheduling event, it estimates projected prefill and decode completion times for candidate placements, computes the standalone execution horizon $H_w$ used as the workflow-level SLO target, and provides the projected ratios used by the scheduler to rank ready calls. Service times are obtained from a roofline-style latency model~\cite{williams2009roofline,cao2025moe_lightning}.

\section{Evaluation}
\label{sec:evaluation}

This section evaluates the performance of \sys under P-D disaggregated serving of agentic workflows. We begin by asking the following questions:
\begin{itemize}
    \item \textbf{(\underline{Q1})} \textit{How does the end-to-end SLO attainment of \sys compare to that of the state-of-the-art LLM serving systems on heterogeneous clusters?}
    \item \textbf{(\underline{Q2})} \textit{How does the end-to-end SLO attainment of \sys compare to that of the state-of-the-art LLM serving systems on homogeneous clusters?}
    \item \textbf{(\underline{Q3})} \textit{How robust is \sys to the estimation error of serving latencies?}
    \item \textbf{(\underline{Q4})} \textit{What scheduling overhead does \sys introduce?}
\end{itemize}

\vspace{-0.5em}
\subsection{Experimental Setup}
\label{subsec:eval-setup}

\vspace{0.5em}
\noindent\textbf{Workloads.}
We evaluate \sys over four representative agentic workloads with multiple LLM calls. Each trace is represented as an online-revealed workflow DAG whose nodes are LLM calls annotated with input length, output length, parent dependencies, and workflow identity. Analogous to Autellix~\cite{luo2025autellix}, we construct traces from the following representative agentic sources:
\begin{itemize}
    \item \textbf{ShareGPT.} This trace contains $100$ conversational-chain workflows arriving at $10$ workflows/s, based on ShareGPT-style conversations used by Vicuna~\cite{chiang2023vicuna}. It primarily stresses sequential workflow progress and accumulated end-to-end chain latency.
    \vspace{0.15em}
    \item \textbf{BFCL-v3.} This trace contains $400$ function-calling workflows arriving at $40$ workflows/s, based on BFCL-v3~\cite{patil2025bfcl}. It represents tool-heavy workloads with frequent online call revelation.
    \vspace{0.15em}
    \item \textbf{LATS.} This trace contains $100$ tree/search-style workflows arriving at $40$ workflows/s. It is derived by running LATS-style Monte Carlo tree search~\cite{zhou2024lats} on HotpotQA~\cite{yang2018hotpotqa}, and stresses reasoning workflows with bursty fan-out, where completing one parent call may reveal multiple downstream calls.
    \vspace{0.15em}
    \item \textbf{Mixed.} This trace interleaves ShareGPT, BFCL-v3, and LATS workflows, with $100$ workflows arriving at $10$ workflows/s. It evaluates scheduler robustness when different agentic workflows share the same P-D serving cluster.
\end{itemize}

\vspace{0.5em}
\noindent\textbf{Models.}
We evaluate on \textsc{Llama3.1-70B} and \textsc{Qwen3-235B-A22B}. The \textsc{Llama3.1-70B} experiments use tensor parallelism degree 4 for A100, H100, and H200 GPUs. The \textsc{Qwen3-235B-A22B} experiments use tensor parallelism degree 8 for A100 and H100, and tensor parallelism degree 4 for H200.

\vspace{0.5em}
\noindent\textbf{Heterogeneous runtime.}
Our main experiments use heterogeneous A100/H100/H200 P-D clusters. We evaluate two heterogeneous configurations:
\begin{itemize}
    \item \underline{Hetero-1}: This setup contains 8 prefill instances and 8 decode instances, with each pool composed of 2 A100, 3 H100, and 3 H200 model instances.
    \vspace{0.15em}
    \item \underline{Hetero-2}: This setup contains 10 prefill instances and 10 decode instances, with each pool composed of 3 A100, 4 H100, and 3 H200 model instances.
\end{itemize}

\vspace{0.5em}
\noindent\textbf{Homogeneous runtime.}
To isolate the benefit of workflow-aware scheduling without hardware heterogeneity, we report a single homogeneous 4P+4D setting for each model family. The Llama homogeneous experiment uses four H200 TP4 prefill instances and four H200 TP4 decode instances. The Qwen homogeneous experiment uses four A100 TP8 prefill instances and four A100 TP8 decode instances.

\vspace{-0.5em}
\subsection{Baselines}
\label{subsec:eval-baselines}

We compare \sys against three strong baselines.

\begin{itemize}
    \item \textbf{SGLang-FCFS.} SGLang~\cite{zheng2024sglang} combines continuous batching with RadixAttention to efficiently manage KV cache across requests. Its default scheduling policy is FCFS. We implement a workflow-level FCFS scheduler with load-balanced dispatching.
    \item \textbf{SGLang-LLF.} On top of SGLang, we implement a workflow-level least-laxity-first scheduler~\cite{stankovic1998scheduling} that prioritizes workflows with smaller remaining slack relative to deadlines.
    \item \textbf{Autellix-ATLAS.} An adaptation of ATLAS from \emph{Autellix}~\cite{luo2025autellix}, representing a state-of-the-art program-aware attained-service baseline.
\end{itemize}

\vspace{-0.5em}
\subsection{Evaluation Metrics}
\label{subsec:eval-metrics}

\vspace{0.5em}
\noindent\textbf{SLO scaling metrics.}
Following standard practice in LLM serving evaluations~\cite{li2023alpaserve,zhong2024distserve}, we evaluate system performance based on workflow-level SLO attainment. For each workflow, we first collect its standalone execution horizon $H_w$ in an exclusive environment on the same serving cluster. During evaluation, we set the SLO by multiplying this standalone horizon with a scale factor $\alpha$, i.e., the workflow will be violating SLO if $C_w \ge \alpha H_w$. Smaller $\alpha$ corresponds to a tighter SLO. We report two tail metrics: \textbf{Req95} and \textbf{Req99}, which represent the minimum SLO scale $\alpha$ such that at least $95\%$ and $99\%$ of workflows complete within $\alpha H_w$, respectively. Lower Req95 and Req99 mean that the scheduler can satisfy the same workflow-level attainment target under tighter end-to-end SLOs. We also visualize full SLO-attainment curves by sweeping $\alpha$.

\vspace{0.5em}
\noindent\textbf{Scheduler overhead.}
We report total scheduler time and average scheduler time per invocation. Since scheduling is asynchronous, this overhead affects quality indirectly: a slower scheduler leaves more calls executing under fallback plans or stale assignments.

\begin{figure*}[t]
\centering
\includegraphics[width=\textwidth]{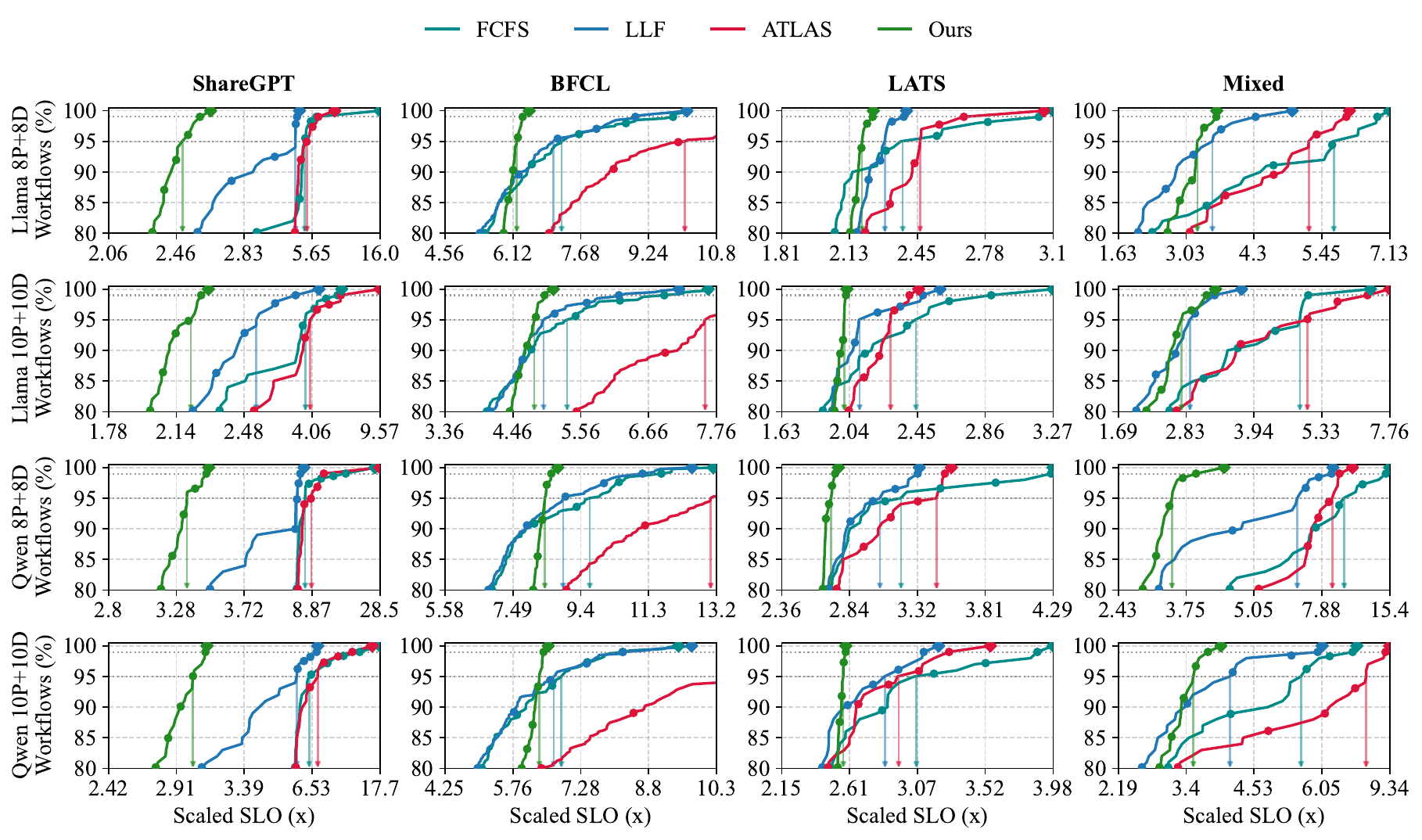}
\vspace{-1.0em}
\caption{SLO-attainment curves on heterogeneous A100/H100/H200 clusters. Each subplot sweeps the SLO scale $\alpha$ and reports the fraction of workflows satisfying $C_w \le \alpha H_w$. Higher curves to the left indicate stronger performance. \sys consistently reaches high attainment at smaller SLO scales, especially on Qwen and mixed/LATS-style workloads.}
\label{fig:hetero-slo-grid}
\end{figure*}

\subsection{End-to-End Evaluation with Heterogeneity  \textbf{(\underline{Q1})}}
\label{subsec:eval-hetero}

\vspace{0.5em}
\noindent\textbf{Overall comparison.}
Table~\ref{tab:hetero-main} summarizes the heterogeneous results. Each row averages over the four workload traces, and the baseline column reports the strongest baseline for each trace before averaging. Across all heterogeneous model/cluster settings, \sys reduces the scale factor required to finish 95\% of workflows by 13.0\% on average and the scale factor required to finish 99\% of workflows by 24.5\% on average. The improvement is consistent across both models and both cluster sizes: even in the weakest averaged case, Llama on Hetero-2, \sys still reduces Req95 by 6.1\% and Req99 by 13.9\% over the strongest baseline.

\vspace{0.25em}
The largest average gains appear on Qwen, where the heterogeneous deployment creates a stronger placement problem. On Qwen Hetero-1, \sys lowers the average Req95 from 5.60 to 4.48 and Req99 from 7.22 to 4.69, corresponding to 21.1\% and 33.1\% reductions. On Qwen Hetero-2, \sys lowers the average Req95 from 4.42 to 3.87 and Req99 from 6.11 to 3.98, corresponding to 13.5\% and 32.7\% reductions. These results indicate that the benefit of \sys becomes more pronounced when workflow-level pressure interacts with expensive model execution and heterogeneous prefill/decode assignment decisions.

\begin{table}[t]
  \centering
  \caption{End-to-end results on heterogeneous A100/H100/H200 clusters. Values are averaged across ShareGPT, BFCL-v3, LATS, and Mixed. Each cell reports Req95/Req99. Each baseline column uses the strongest baseline for the corresponding trace. Lower is better.}
  \label{tab:hetero-main}
  \vspace{-0.5em}
  \resizebox{0.6\columnwidth}{!}{%
  \begin{tabular}{l|l|c|c|c}
  \hline
  \textbf{Model} & \textbf{Cluster} & \textbf{\sys} & \textbf{Best baseline} & \textbf{Reduction} \\
  \hline
  Llama & Hetero-1 & 3.54 / 3.71 & 4.00 / 4.75 & 11.2\% / 18.2\% \\
  \hline
  Llama & Hetero-2 & 2.95 / 3.11 & 3.12 / 3.65 & 6.1\% / 13.9\% \\
  \hline
  Qwen  & Hetero-1 & 4.48 / 4.69 & 5.60 / 7.22 & 21.1\% / 33.1\% \\
  \hline
  Qwen  & Hetero-2 & 3.87 / 3.98 & 4.42 / 6.11 & 13.5\% / 32.7\% \\
  \hline
  \end{tabular}%
  }
  \vspace{-0.5em}
\end{table}

\begin{table}[t]
  \centering
  \caption{Detailed heterogeneous results for Qwen on Hetero-1. Each cell reports Req95/Req99. This setting illustrates \sys's behavior under strong heterogeneous queueing pressure. Lower is better.}
  \label{tab:hetero-qwen-detail}
  \vspace{-0.5em}
  \resizebox{0.65\columnwidth}{!}{%
  \begin{tabular}{l|c|c|c}
  \hline
  \textbf{Trace} & \textbf{\sys} & \textbf{Best baseline} & \textbf{Reduction} \\
  \hline
  ShareGPT & 3.36 / 3.50 & 4.58 / 5.45 (Workflow-LLF) & 26.7\% / 35.9\% \\
  \hline
  BFCL-v3  & 8.39 / 8.57 & 8.90 / 11.13 (Workflow-LLF) & 5.7\% / 23.0\% \\
  \hline
  LATS     & 2.71 / 2.74 & 3.06 / 3.33 (Workflow-LLF) & 11.3\% / 17.7\% \\
  \hline
  Mixed    & 3.48 / 3.94 & 5.85 / 8.96 (Workflow-LLF) & 40.6\% / 56.0\% \\
  \hline
  \end{tabular}%
  }
  \vspace{-0.5em}
\end{table}

\vspace{0.5em}
\noindent\textbf{Where does the gain come from?}
Figure~\ref{fig:hetero-slo-grid} shows the full SLO-attainment curves and Table~\ref{tab:hetero-qwen-detail} shows the per-trace results for Qwen on Hetero-1, the heterogeneous Qwen setting where \sys obtains the strongest average improvement. \sys improves Req95 and Req99 on all traces. The most significant tail-latency reduction appears on Mixed: \sys reduces Req99 from 8.96 to 3.94, a 56.0\% reduction over Workflow-LLF. It also reduces Req99 on ShareGPT from 5.45 to 3.50, a 35.9\% reduction over Workflow-LLF, and reduces Req99 on BFCL-v3 from 11.13 to 8.57, a 23.0\% reduction over Workflow-LLF. LATS also benefits: \sys reduces Req95 from 3.06 to 2.71 and Req99 from 3.33 to 2.74, corresponding to 11.3\% and 17.7\% reductions.

\vspace{0.25em}
The Mixed and ShareGPT results illustrate that workflow-aware placement matters. Mixed workloads combine chain-like, function-calling, and tree-search workflows, creating non-uniform queueing pressure across both P and D stages. ShareGPT workflows are more sequential, so delaying one revealed call can directly delay the entire workflow. Workflow-FCFS preserves arrival order but cannot distinguish workflows whose revealed frontier is close to violating its scaled SLO. Workflow-LLF captures urgency, but it does not jointly evaluate which prefill/decode pair will minimize the projected normalized completion after accounting for heterogeneous service latencies and transfer-induced decode-ready times. Autellix-ATLAS tracks attained service at the program level, which helps fairness, but attained service is not the same as the remaining risk of exceeding a workflow-specific horizon. In contrast, \sys directly estimates the consequence of scheduling each ready call on a concrete prefill/decode pair and prioritizes calls that most threaten workflow-level completion. This explains why the largest gains appear on Mixed and ShareGPT traces, where workflow-level queueing pressure and heterogeneous placement sensitivity are both pronounced.

\vspace{0.25em}
The BFCL-v3 and LATS rows provide complementary context. BFCL-v3 consists of many short function-calling tasks, and Workflow-LLF already approximates much of the urgency signal, so the remaining room for improvement is smaller at Req95. Nevertheless, \sys still reduces Req99 by 23.0\%, showing that workflow-aware prefill/decode placement improves the tail even when the average scheduling problem is relatively easy for strong urgency-based baselines. LATS has tree/search-style online revelation, and \sys improves both Req95 and Req99 by avoiding placements that create downstream queueing pressure after new branches are revealed.

\begin{figure*}[t]
\centering
\includegraphics[width=\linewidth]{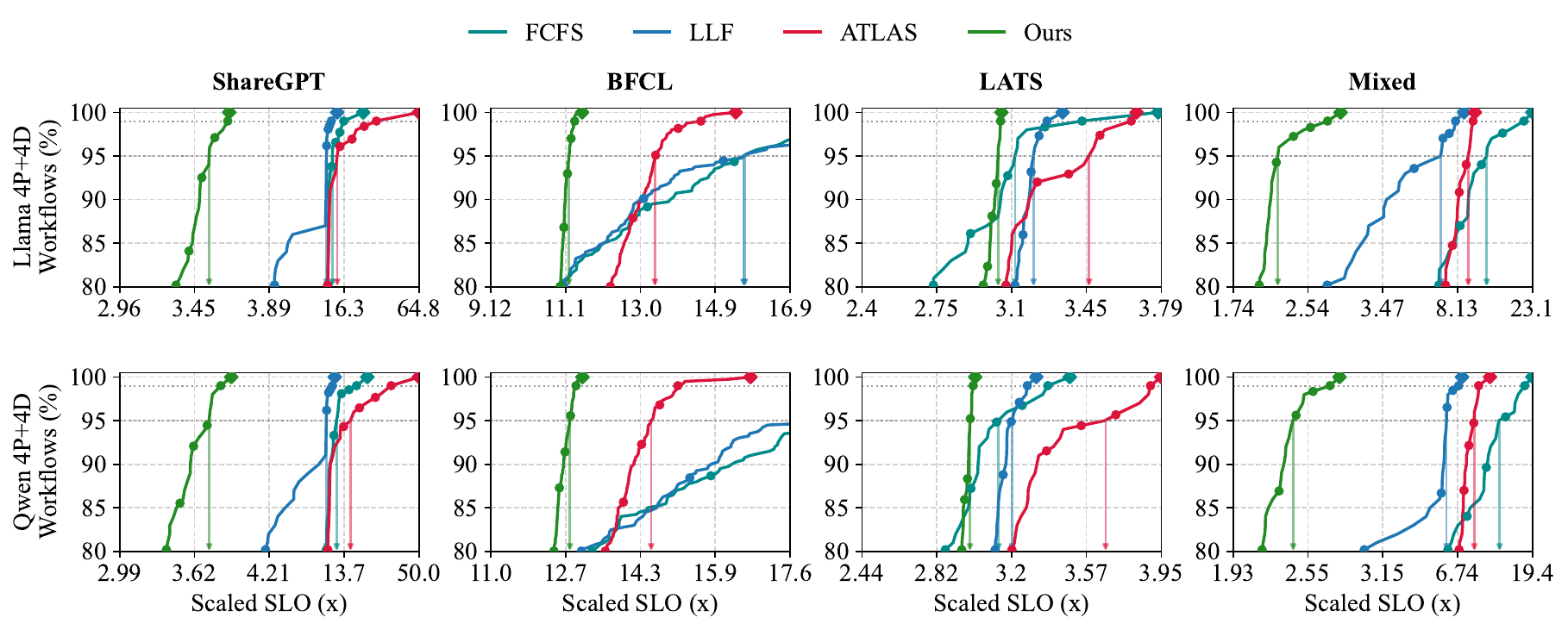}
\vspace{-1.0em}
\caption{SLO-attainment curves on homogeneous 4P+4D clusters. Each subplot sweeps the SLO scale $\alpha$ and reports the fraction of workflows satisfying $C_w \le \alpha H_w$. Even without heterogeneous placement choices, \sys reaches high attainment at smaller scales by improving workflow-aware queue ordering.}
\label{fig:homo-slo-grid}
\end{figure*}

\subsection{Evaluation on Homogeneous Clusters \textbf{(\underline{Q2})}}
\label{subsec:eval-homo}

\vspace{0.5em}
\noindent\textbf{Homogeneous results.}
Figure~\ref{fig:homo-slo-grid} shows the SLO-attainment curves for the homogeneous setting, and Table~\ref{tab:homo-main} summarizes the corresponding tail metrics. \sys improves Req95 and Req99 for both model families. These results show that \sys reduces Req95 by $24.8\%$ and Req99 by $36.8\%$ on the Llama model executed on the H200 GPU, and reduces Req95 by $23.1\%$ and Req99 by $33.9\%$ on the Qwen model executed on the A100 GPUs.

\vspace{0.25em}
The homogeneous results show that \sys does not rely solely on hardware heterogeneity. Even when all instances in a cluster share the same model configuration, workflow-aware scheduling still reduces tail SLO scales by prioritizing workflows that are closest to their current SLO target and by avoiding poor local queue ordering. In particular, the homogeneous curves in Figure~\ref{fig:homo-slo-grid} show that \sys moves the attainment curve left on sequential and mixed workloads, where delaying one call can directly delay the entire workflow. The relative gains are smaller on short structured workloads such as BFCL-v3, where the strongest baselines already keep small tasks moving.

\begin{table}[t]
  \centering
  \caption{Homogeneous 4P+4D results. Values are averaged across ShareGPT, BFCL-v3, LATS, and Mixed. Each cell reports Req95/Req99. Lower is better.}
  \label{tab:homo-main}
  \vspace{-0.5em}
  \resizebox{0.7\columnwidth}{!}{%
  \begin{tabular}{l|l|c|c|c}
  \hline
  \textbf{Model} & \textbf{Cluster} & \textbf{\sys} & \textbf{Best baseline} & \textbf{Red.} \\
  \hline
  Llama & H200 4P+4D & 4.99 / 5.19 & 6.52 / 8.37 & 24.8\% / 36.8\% \\
  \hline
  Qwen  & A100 4P+4D & 5.48 / 5.62 & 6.89 / 8.33 & 23.1\% / 33.9\% \\
  \hline
  \end{tabular}%
  }
  \vspace{-0.5em}
\end{table}

\subsection{Robustness to Estimation Error \textbf{(\underline{Q3})}}
\label{subsec:eval-robustness}

\vspace{0.5em}
\noindent\textbf{Setup.}
We evaluate robustness on the main heterogeneous Hetero-1 setting by injecting deterministic multiplicative error into the scheduler-visible prefill and decode estimates. The actual service durations are unchanged. We sweep estimation error levels of 0\%, 10\%, 20\%, and 30\% for both Llama and Qwen experiments.

\vspace{0.5em}
\noindent\textbf{Results.}
Table~\ref{tab:estimation-error} reports the average degradation in Req95 and Req99 relative to the no-error case. \sys is robust to moderate prediction errors. For Llama, the average Req99 degradation remains within 1.5\% even at 30\% estimation error. Qwen shows larger Req99 sensitivity, especially at 10\% noise, but the average degradation remains bounded and non-monotonic because the scheduler is still driven by workflow-level priorities rather than exact per-call durations.

\begin{table}[t]
  \centering
  \caption{Robustness to estimation error on Hetero-1. Entries report the average percentage
  degradation relative to the 0\% error case across four traces. Lower is better.}
  \label{tab:estimation-error}
  \vspace{-0.5em}
  \resizebox{0.4\linewidth}{!}{%
  \begin{tabular}{l|c|c|c}
  \hline
  \textbf{Model} & \textbf{Error} & \textbf{Req95 deg.} & \textbf{Req99 deg.} \\
  \hline
  Llama & 10\% & 0.9\% & 1.5\% \\
  \hline
  Llama & 20\% & -0.5\% & 0.8\% \\
  \hline
  Llama & 30\% & 0.7\% & 1.5\% \\
  \hline
  Qwen  & 10\% & -1.9\% & 9.5\% \\
  \hline
  Qwen  & 20\% & -1.1\% & 7.0\% \\
  \hline
  Qwen  & 30\% & -1.4\% & 5.4\% \\
  \hline
  \end{tabular}
  }
  \vspace{-0.5em}
  \end{table}

The small degradation indicates that \sys does not require exact service-time prediction to be useful. Estimation error can change the chosen prefill/decode pair for some calls, but the ordering signal is smoothed at the workflow level: the scheduler prioritizes workflows by normalized completion pressure rather than by a single noisy call duration. The negative degradation values in some rows mean that a noisy estimate happened to produce a slightly better greedy ordering than the no-error estimate for that trace; we treat these as near-ties rather than meaningful improvements. Overall, even with 30\% multiplicative estimation error, the scheduler remains stable enough for the workflow-level benefits observed in the main experiments.

\subsection{Scheduler Overhead Study \textbf{(\underline{Q4})}}
\label{subsec:eval-overhead}

\vspace{0.5em}
\noindent\textbf{Asynchronous overhead.}
\sys performs more planning than the workflow-level baselines because it evaluates candidate prefill/decode placements and recomputes priorities for ready calls. However, the asynchronous runtime prevents scheduling from directly blocking already running service. In our heterogeneous overhead screen, \sys requires only millisecond-scale planning: the average scheduling time per invocation is 10.2 ms for Llama Hetero-1, 7.1 ms for Llama Hetero-2, 14.7 ms for Qwen Hetero-1, and 11.5 ms for Qwen Hetero-2. The worst per-invocation average across traces is 23.7 ms. These numbers indicate that the additional workflow-aware planning cost is small relative to end-to-end workflow execution time.

\begin{table}[t]
  \centering
  \caption{Scheduler overhead in heterogeneous settings. Values are averaged across traces for \sys. Ms/inv represents the overhead (ms) per scheduler invocation }
  \label{tab:scheduler-overhead}
  \vspace{-0.5em}
  \resizebox{0.45\linewidth}{!}{%
  \begin{tabular}{l|l|c|c}
  \hline
  \textbf{Model} & \textbf{Cluster} & \textbf{Ms/inv} & \textbf{Total overhead (s)}
  \\
  \hline
  Llama & Hetero-1 & 10.2 & 25.7 \\
  \hline
  Llama & Hetero-2 & 7.1 & 19.4 \\
  \hline
  Qwen  & Hetero-1 & 14.7 & 28.8 \\
  \hline
  Qwen  & Hetero-2 & 11.5 & 23.2 \\
  \hline
  \end{tabular}%
  }
  \vspace{-0.5em}
  \end{table}

\vspace{0.5em}
\noindent\textbf{Impact on serving quality.}
The overhead results should be interpreted together with the end-to-end tables. A purely greedy local policy is cheaper, but it can leave workflows with much larger tail SLO scales. \sys spends a few to tens of milliseconds to evaluate the consequences of placement and ordering decisions, and the asynchronous plan-application rule ensures that this planning cost does not stall calls already in service. Thus, the scheduler trades a small amount of CPU-side planning work for substantially tighter workflow-level SLO scaling, especially in heterogeneous settings where a wrong decode placement or a stale local queue order can dominate the tail.

\section{Conclusion}
\label{sec:conclusion}

We present \sys, a workflow- \& heterogeneity-aware scheduler for serving agentic LLM requests on the P-D disaggregated paradigm. \sys targets end-to-end workflow SLO attainment by maintaining an online planning horizon for each partially revealed workflow and jointly plans prefill and decode assignment, accounts for heterogeneous serving latencies. Our evaluation shows that \sys improves tail scaled-SLO attainment over program-aware baselines in the heterogeneous setting, and that the benefit also carries over to homogeneous H200 clusters. These results suggest that workflow-level objectives should become first-class scheduling targets as LLM applications evolve from isolated requests into structured, multi-step agentic systems.


\bibliographystyle{plainnat}
\bibliography{reference}

@article{jiang2026oserve,
  title={OServe: Accelerating LLM Serving via Spatial-Temporal Workload Orchestration},
  author={Jiang, Youhe and Fu, Fangcheng and Wang, Taiyi and He, Guoliang and Yoneki, Eiko},
  journal={arXiv preprint arXiv:2602.12151},
  year={2026}
}

@article{zhang2025efficient,
  title={Efficient mixed-precision large language model inference with turbomind},
  author={Zhang, Li and Jiang, Youhe and He, Guoliang and Chen, Xin and Lv, Han and Yao, Qian and Fu, Fangcheng and Chen, Kai},
  journal={arXiv preprint arXiv:2508.15601},
  year={2025}
}

@article{jiang2025cascadia,
  title={Cascadia: An Efficient Cascade Serving System for Large Language Models},
  author={Jiang, Youhe and Fu, Fangcheng and Zhao, Wanru and Rabanser, Stephan and Zhang, Jintao and Lane, Nicholas D and Yuan, Binhang},
  journal={arXiv preprint arXiv:2506.04203},
  year={2025}
}

@article{miao2022galvatron,
  title={Galvatron: Efficient transformer training over multiple gpus using automatic parallelism},
  author={Miao, Xupeng and Wang, Yujie and Jiang, Youhe and Shi, Chunan and Nie, Xiaonan and Zhang, Hailin and Cui, Bin},
  journal={arXiv preprint arXiv:2211.13878},
  year={2022}
}

@article{yan2025fsa,
  title={Fsa: An alternative efficient implementation of native sparse attention kernel},
  author={Yan, Ran and Jiang, Youhe and Chen, Zhuoming and Mai, Haohui and Chen, Beidi and Yuan, Binhang},
  journal={arXiv preprint arXiv:2508.18224},
  year={2025}
}

@article{jiang2022osdp,
  title={Osdp: Optimal sharded data parallel for distributed deep learning},
  author={Jiang, Youhe and Fu, Fangcheng and Miao, Xupeng and Nie, Xiaonan and Cui, Bin},
  journal={arXiv preprint arXiv:2209.13258},
  year={2022}
}

@article{wang2024improving,
  title={Improving automatic parallel training via balanced memory workload optimization},
  author={Wang, Yujie and Jiang, Youhe and Miao, Xupeng and Fu, Fangcheng and Zhu, Shenhan and Nie, Xiaonan and Tu, Yaofeng and Cui, Bin},
  journal={IEEE Transactions on Knowledge and Data Engineering},
  volume={36},
  number={8},
  pages={3906--3920},
  year={2024},
  publisher={IEEE}
}

@article{he2026efficient,
  title={Efficient Pre-Training of LLMs via Topology-Aware Communication Alignment on More Than 9600 GPUs},
  author={He, Guoliang and Jiang, Youhe and Xiao, Wencong and Kaihua, Jiang and Wang, Shuguang and Wang, Jun and Zixian, Du and Jiang, Zhuo and Zhang, Xinlei and Yuan, Binhang and others},
  journal={Advances in Neural Information Processing Systems},
  volume={38},
  pages={147100--147126},
  year={2026}
}

@article{he2026efficient2,
  title={Efficient Multi-round LLM Inference over Disaggregated Serving},
  author={He, Wenhao and Jiang, Youhe and Zhao, Penghao and Xu, Quanqing and Yoneki, Eiko and Cui, Bin and Fu, Fangcheng},
  journal={arXiv preprint arXiv:2602.14516},
  year={2026}
}

@article{jiang2025demystifying,
  title={Demystifying cost-efficiency in llm serving over heterogeneous gpus},
  author={Jiang, Youhe and Fu, Fangcheng and Yao, Xiaozhe and He, Guoliang and Miao, Xupeng and Klimovic, Ana and Cui, Bin and Yuan, Binhang and Yoneki, Eiko},
  journal={arXiv preprint arXiv:2502.00722},
  year={2025}
}

@article{tong2025parallax,
  title={Parallax: Efficient llm inference service over decentralized environment},
  author={Tong, Chris and Jiang, Youhe and Chen, Gufeng and Zhao, Tianyi and Lu, Sibian and Qu, Wenjie and Yang, Eric and Ai, Lynn and Yuan, Binhang},
  journal={arXiv preprint arXiv:2509.26182},
  year={2025}
}

@article{jiang2026autopoiesis,
  title={Autopoiesis: A Self-Evolving System Paradigm for LLM Serving Under Runtime Dynamics},
  author={Jiang, Youhe and Yan, Ran and Peng, You and Li, Wenshuang and Wang, Taiyi and Fu, Fangcheng and Yuan, Binhang},
  journal={arXiv preprint arXiv:2604.07144},
  year={2026}
}

@article{peng2025hexgen,
  title={HEXGEN-FLOW: Optimizing LLM Inference Request Scheduling for Agentic Text-to-SQL},
  author={Peng, You and Jiang, Youhe and Jiang, Wenqi and Wang, Chen and Yuan, Binhang},
  journal={arXiv preprint arXiv:2505.05286},
  year={2025}
}

@article{yan2025areal,
  title={AReaL-Hex: Accommodating Asynchronous RL Training over Heterogeneous GPUs},
  author={Yan, Ran and Jiang, Youhe and Wu, Tianyuan and Gao, Jiaxuan and Mei, Zhiyu and Fu, Wei and Mai, Haohui and Wang, Wei and Wu, Yi and Yuan, Binhang},
  journal={arXiv preprint arXiv:2511.00796},
  year={2025}
}

@article{jiang2026boute,
  title={BOute: Cost-Efficient LLM Serving with Heterogeneous LLMs and GPUs via Multi-Objective Bayesian Optimization},
  author={Jiang, Youhe and Fu, Fangcheng and Yoneki, Eiko},
  journal={arXiv preprint arXiv:2602.10729},
  year={2026}
}

@inproceedings{yu2022orca,
  title     = {Orca: A Distributed Serving System for Transformer-Based Generative Models},
  author    = {Gyeong-In Yu and Joo Seong Jeong and Geon-Woo Kim and Soojeong Kim and Byung-Gon Chun},
  booktitle = {16th USENIX Symposium on Operating Systems Design and Implementation (OSDI 22)},
  pages     = {521--538},
  year      = {2022},
  publisher = {USENIX Association},
  address   = {Carlsbad, CA},
  month     = jul,
  isbn      = {978-1-939133-28-1},
  url       = {https://www.usenix.org/conference/osdi22/presentation/yu}
}

@article{jiang2025hexgen,
  title={Hexgen-2: Disaggregated generative inference of llms in heterogeneous environment},
  author={Jiang, Youhe and Yan, Ran and Yuan, Binhang},
  journal={arXiv preprint arXiv:2502.07903},
  year={2025}
}

@article{jiang2023hexgen,
  title={Hexgen: Generative inference of large language model over heterogeneous environment},
  author={Jiang, Youhe and Yan, Ran and Yao, Xiaozhe and Zhou, Yang and Chen, Beidi and Yuan, Binhang},
  journal={arXiv preprint arXiv:2311.11514},
  year={2023}
}

@inproceedings{mao2025skyserve,
  title={Skyserve: Serving ai models across regions and clouds with spot instances},
  author={Mao, Ziming and Xia, Tian and Wu, Zhanghao and Chiang, Wei-Lin and Griggs, Tyler and Bhardwaj, Romil and Yang, Zongheng and Shenker, Scott and Stoica, Ion},
  booktitle={Proceedings of the Twentieth European Conference on Computer Systems},
  pages={159--175},
  year={2025}
}

@article{jiang2025thunderserve,
  title={Thunderserve: High-performance and cost-efficient llm serving in cloud environments},
  author={Jiang, Youhe and Fu, Fangcheng and Yao, Xiaozhe and Wang, Taiyi and Cui, Bin and Klimovic, Ana and Yoneki, Eiko},
  journal={Proceedings of Machine Learning and Systems},
  volume={7},
  year={2025}
}

@inproceedings{patel2024splitwise,
  title={Splitwise: Efficient generative llm inference using phase splitting},
  author={Patel, Pratyush and Choukse, Esha and Zhang, Chaojie and Shah, Aashaka and Goiri, {\'I}{\~n}igo and Maleki, Saeed and Bianchini, Ricardo},
  booktitle={2024 ACM/IEEE 51st Annual International Symposium on Computer Architecture (ISCA)},
  pages={118--132},
  year={2024},
  organization={IEEE}
}

@inproceedings{kwon2023vllm,
  title     = {Efficient Memory Management for Large Language Model Serving with {PagedAttention}},
  author    = {Woosuk Kwon and Zhuohan Li and Siyuan Zhuang and Ying Sheng and Lianmin Zheng and Cody Hao Yu and Joseph E. Gonzalez and Hao Zhang and Ion Stoica},
  booktitle = {Proceedings of the ACM SIGOPS 29th Symposium on Operating Systems Principles},
  pages     = {611--626},
  year      = {2023},
  publisher = {Association for Computing Machinery},
  address   = {New York, NY, USA},
  doi       = {10.1145/3600006.3613165}
}

@inproceedings{zheng2024sglang,
  title     = {{SGLang}: Efficient Execution of Structured Language Model Programs},
  author    = {Lianmin Zheng and Liangsheng Yin and Zhiqiang Xie and Chuyue Sun and Jeff Huang and Cody Hao Yu and Shiyi Cao and Christos Kozyrakis and Ion Stoica and Joseph E. Gonzalez and Clark W. Barrett and Ying Sheng},
  booktitle = {Advances in Neural Information Processing Systems},
  volume    = {37},
  year      = {2024},
  publisher = {Curran Associates, Inc.},
  address   = {Red Hook, NY, USA},
  url       = {https://papers.nips.cc/paper_files/paper/2024/hash/724be4472168f31ba1c9ac630f15dec8-Abstract-Conference.html}
}

@inproceedings{agrawal2024sarathiserve,
  title     = {Taming Throughput-Latency Trade-off in {LLM} Inference with {Sarathi-Serve}},
  author    = {Amey Agrawal and Nitin Kedia and Ashish Panwar and Jayashree Mohan and Nipun Kwatra and Bhargav S. Gulavani and Alexey Tumanov and Ramachandran Ramjee},
  booktitle = {18th USENIX Symposium on Operating Systems Design and Implementation (OSDI 24)},
  year      = {2024},
  publisher = {USENIX Association},
  address   = {Santa Clara, CA},
  month     = jul,
  isbn      = {978-1-939133-40-3},
  url       = {https://www.usenix.org/conference/osdi24/presentation/agrawal}
}

@inproceedings{sun2024llumnix,
  title     = {Llumnix: Dynamic Scheduling for Large Language Model Serving},
  author    = {Biao Sun and Ziming Huang and Hanyu Zhao and Wencong Xiao and Xinyi Zhang and Yong Li and Wei Lin},
  booktitle = {18th USENIX Symposium on Operating Systems Design and Implementation (OSDI 24)},
  pages     = {173--191},
  year      = {2024},
  publisher = {USENIX Association},
  address   = {Santa Clara, CA},
  month     = jul,
  isbn      = {978-1-939133-40-3},
  url       = {https://www.usenix.org/conference/osdi24/presentation/sun-biao}
}

@inproceedings{fu2024serverlessllm,
  title     = {{ServerlessLLM}: Low-Latency Serverless Inference for Large Language Models},
  author    = {Yao Fu and Leyang Xue and Yeqi Huang and Andrei-Octavian Brabete and Dmitrii Ustiugov and Yuvraj Patel and Luo Mai},
  booktitle = {18th USENIX Symposium on Operating Systems Design and Implementation (OSDI 24)},
  pages     = {135--153},
  year      = {2024},
  publisher = {USENIX Association},
  address   = {Santa Clara, CA},
  month     = jul,
  isbn      = {978-1-939133-40-3},
  url       = {https://www.usenix.org/conference/osdi24/presentation/fu}
}

@inproceedings{zhong2024distserve,
  title     = {DistServe: Disaggregating Prefill and Decoding for Goodput-optimized Large Language Model Serving},
  author    = {Yinmin Zhong and Shengyu Liu and Junda Chen and Jianbo Hu and Yibo Zhu and Xuanzhe Liu and Xin Jin and Hao Zhang},
  booktitle = {18th USENIX Symposium on Operating Systems Design and Implementation (OSDI 24)},
  pages     = {193--210},
  year      = {2024},
  publisher = {USENIX Association},
  address   = {Santa Clara, CA},
  month     = jul,
  isbn      = {978-1-939133-40-3},
  url       = {https://www.usenix.org/conference/osdi24/presentation/zhong-yinmin}
}

@misc{jin2024pdserve,
  title        = {{P/D-Serve}: Serving Disaggregated Large Language Model at Scale},
  author       = {Yibo Jin and Tao Wang and Huimin Lin and Mingyang Song and Peiyang Li and Yipeng Ma and Yicheng Shan and Zhengfan Yuan and Cailong Li and Yajing Sun and Tiandeng Wu and Xing Chu and Ruizhi Huan and Li Ma and Xiao You and Wenting Zhou and Yunpeng Ye and Wen Liu and Xiangkun Xu and Yongsheng Zhang and Tiantian Dong and Jiawei Zhu and Zhe Wang and Xijian Ju and Jianxun Song and Haoliang Cheng and Xiaojing Li and Jiandong Ding and Hefei Guo and Zhengyong Zhang},
  year         = {2024},
  eprint       = {2408.08147},
  archivePrefix= {arXiv},
  primaryClass = {cs.DC},
  doi          = {10.48550/arXiv.2408.08147}
}

@inproceedings{lin2024parrot,
  title     = {Parrot: Efficient Serving of {LLM}-based Applications with Semantic Variable},
  author    = {Chaofan Lin and Zhenhua Han and Chengruidong Zhang and Yuqing Yang and Fan Yang and Chen Chen and Lili Qiu},
  booktitle = {18th USENIX Symposium on Operating Systems Design and Implementation (OSDI 24)},
  pages     = {929--945},
  year      = {2024},
  publisher = {USENIX Association},
  address   = {Santa Clara, CA},
  month     = jul,
  isbn      = {978-1-939133-40-3},
  url       = {https://www.usenix.org/conference/osdi24/presentation/lin-chaofan}
}

@article{liu2026hermes,
  title   = {Hermes: Efficient Serving of {LLM} Applications with Probabilistic Demand Modeling},
  author  = {Yifei Liu and Zuo Gan and Zhenghao Gan and Weiye Wang and Chen Chen and Yizhou Shan and Xusheng Chen and Zhenhua Han and Yifei Zhu and Shixuan Sun and Minyi Guo},
  journal = {ACM Transactions on Architecture and Code Optimization},
  year    = {2026},
  doi     = {10.1145/3803390}
}

@misc{luo2025autellix,
  title        = {Autellix: An Efficient Serving Engine for {LLM} Agents as General Programs},
  author       = {Michael Luo and Xiaoxiang Shi and Colin Cai and Tianjun Zhang and Justin Wong and Yichuan Wang and Chi Wang and Yanping Huang and Zhifeng Chen and Joseph E. Gonzalez and Ion Stoica},
  year         = {2025},
  eprint       = {2502.13965},
  archivePrefix= {arXiv},
  primaryClass = {cs.DC},
  doi          = {10.48550/arXiv.2502.13965}
}

@misc{li2025continuum,
  title        = {Continuum: Efficient and Robust Multi-Turn {LLM} Agent Scheduling with {KV} Cache Time-to-Live},
  author       = {Hanchen Li and Qiuyang Mang and Runyuan He and Qizheng Zhang and Huanzhi Mao and Xiaokun Chen and Alvin Cheung and Joseph E. Gonzalez and Ion Stoica},
  year         = {2025},
  eprint       = {2511.02230},
  archivePrefix= {arXiv},
  primaryClass = {cs.DC},
  doi          = {10.48550/arXiv.2511.02230}
}

@inproceedings{patke2024qlm,
  title     = {Queue Management for {SLO}-Oriented Large Language Model Serving},
  author    = {Archit Patke and Dhemath Reddy and Saurabh Jha and Haoran Qiu and Christian Pinto and Chandra Narayanaswami and Zbigniew T. Kalbarczyk and Ravishankar K. Iyer},
  booktitle = {Proceedings of the 2024 ACM Symposium on Cloud Computing},
  pages     = {18--35},
  year      = {2024},
  publisher = {Association for Computing Machinery},
  address   = {New York, NY, USA},
  doi       = {10.1145/3698038.3698523}
}

@inproceedings{qiu2024ssjf,
  title     = {Efficient Interactive {LLM} Serving with Proxy Model-based Sequence Length Prediction},
  author    = {Haoran Qiu and Weichao Mao and Archit Patke and Shengkun Cui and Saurabh Jha and Chen Wang and Hubertus Franke and Zbigniew T. Kalbarczyk and Tamer Ba{s}\c{a}r and Ravishankar K. Iyer},
  booktitle = {Proceedings of the 5th International Workshop on Cloud Intelligence / {AIOps} at {ASPLOS} 2024},
  year      = {2024},
  publisher = {Association for Computing Machinery},
  address   = {New York, NY, USA},
  url       = {https://haoran-qiu.com/publication/asplos-2024/}
}

@misc{fu2024ltr,
  title        = {Efficient {LLM} Scheduling by Learning to Rank},
  author       = {Yichao Fu and Siqi Zhu and Runlong Su and Aurick Qiao and Ion Stoica and Hao Zhang},
  year         = {2024},
  eprint       = {2408.15792},
  archivePrefix= {arXiv},
  primaryClass = {cs.LG},
  doi          = {10.48550/arXiv.2408.15792}
}

@misc{shahout2024dontstop,
  title        = {Don't Stop Me Now: Embedding Based Scheduling for {LLM}s},
  author       = {Rana Shahout and Eran Malach and Chunwei Liu and Weifan Jiang and Minlan Yu and Michael Mitzenmacher},
  year         = {2024},
  eprint       = {2410.01035},
  archivePrefix= {arXiv},
  primaryClass = {cs.LG},
  doi          = {10.48550/arXiv.2410.01035}
}

@misc{wu2023fastserve,
  title        = {Fast Distributed Inference Serving for Large Language Models},
  author       = {Bingyang Wu and Yinmin Zhong and Zili Zhang and Gang Huang and Xuanzhe Liu and Xin Jin},
  year         = {2023},
  eprint       = {2305.05920},
  archivePrefix= {arXiv},
  primaryClass = {cs.DC},
  doi          = {10.48550/arXiv.2305.05920}
}

@inproceedings{yao2023react,
  title     = {{ReAct}: Synergizing Reasoning and Acting in Language Models},
  author    = {Shunyu Yao and Jeffrey Zhao and Dian Yu and Nan Du and Izhak Shafran and Karthik R. Narasimhan and Yuan Cao},
  booktitle = {The Eleventh International Conference on Learning Representations},
  year      = {2023},
  url       = {https://openreview.net/forum?id=WE_vluYUL-X}
}

@inproceedings{zhou2024lats,
  title     = {Language Agent Tree Search Unifies Reasoning, Acting, and Planning in Language Models},
  author    = {Andy Zhou and Kai Yan and Michal Shlapentokh-Rothman and Haohan Wang and Yu-Xiong Wang},
  booktitle = {Proceedings of the 41st International Conference on Machine Learning},
  series    = {Proceedings of Machine Learning Research},
  volume    = {235},
  pages     = {62138--62160},
  year      = {2024},
  publisher = {PMLR},
  url       = {https://proceedings.mlr.press/v235/zhou24r.html}
}

@inproceedings{patil2025bfcl,
  title     = {The Berkeley Function Calling Leaderboard ({BFCL}): From Tool Use to Agentic Evaluation of Large Language Models},
  author    = {Shishir G. Patil and Huanzhi Mao and Fanjia Yan and Charlie Cheng-Jie Ji and Vishnu Suresh and Ion Stoica and Joseph E. Gonzalez},
  booktitle = {Forty-second International Conference on Machine Learning},
  year      = {2025},
  url       = {https://icml.cc/virtual/2025/poster/46593}
}

@misc{chiang2023vicuna,
  title        = {Vicuna: An Open-Source Chatbot Impressing {GPT}-4 with 90\%* {ChatGPT} Quality},
  author       = {Wei-Lin Chiang and Zhuohan Li and Zi Lin and Ying Sheng and Zhanghao Wu and Hao Zhang and Lianmin Zheng and Siyuan Zhuang and Yonghao Zhuang and Joseph E. Gonzalez and Ion Stoica and Eric P. Xing},
  howpublished = {LMSYS Blog},
  month        = mar,
  year         = {2023},
  url          = {https://lmsys.org/blog/2023-03-30-vicuna/}
}

@article{schrage1968proof,
  title   = {A Proof of the Optimality of the Shortest Remaining Processing Time Discipline},
  author  = {Linus Schrage},
  journal = {Operations Research},
  volume  = {16},
  number  = {3},
  pages   = {687--690},
  year    = {1968}
}

@inproceedings{davis1993slack,
  title     = {Scheduling Slack Time in Fixed Priority Pre-emptive Systems},
  author    = {Robert I. Davis and Ken W. Tindell and Alan Burns},
  booktitle = {Proceedings of the 14th IEEE Real-Time Systems Symposium},
  pages     = {222--231},
  year      = {1993},
  publisher = {IEEE Computer Society},
  address   = {Washington, DC, USA},
  doi       = {10.1109/REAL.1993.393505}
}

@misc{zhao2024blendserve,
  title        = {BlendServe: Optimizing Offline Inference for Auto-regressive Large Models with Resource-aware Batching},
  author       = {Yilong Zhao and Shuo Yang and Kan Zhu and Lianmin Zheng and Baris Kasikci and Yang Zhou and Jiarong Xing and Ion Stoica},
  year         = {2024},
  eprint       = {2411.16102},
  archivePrefix= {arXiv},
  primaryClass = {cs.DC},
  doi          = {10.48550/arXiv.2411.16102}
}

@inproceedings{qiu2024proxy,
  author    = {Haoran Qiu and Weichao Mao and Archit Patke and Shengkun Cui and Saurabh Jha and Chen Wang and Hubertus Franke and Zbigniew T. Kalbarczyk and Tamer Basar and Ravishankar K. Iyer},
  title     = {Efficient Interactive LLM Serving with Proxy Model-based Sequence Length Prediction},
  booktitle = {Proceedings of the 5th International Workshop on Cloud Intelligence / AIOps at ASPLOS 2024 (AIOps 2024)},
  pages     = {1--7},
  year      = {2024},
  address   = {San Diego, CA, USA},
  publisher = {Association for Computing Machinery}
}

@inproceedings{yang2018hotpotqa,
  title     = {{H}otpot{QA}: A Dataset for Diverse, Explainable Multi-hop Question Answering},
  author    = {Yang, Zhilin and Qi, Peng and Zhang, Saizheng and Bengio, Yoshua and Cohen, William and Salakhutdinov, Ruslan and Manning, Christopher D.},
  booktitle = {Proceedings of the 2018 Conference on Empirical Methods in Natural Language Processing},
  pages     = {2369--2380},
  year      = {2018},
  publisher = {Association for Computational Linguistics},
  doi       = {10.18653/v1/D18-1259},
  url       = {https://aclanthology.org/D18-1259/}
}

@inproceedings{li2023alpaserve,
  title     = {{AlpaServe}: Statistical Multiplexing with Model Parallelism for Deep Learning Serving},
  author    = {Zhuohan Li and Lianmin Zheng and Yinmin Zhong and Vincent Liu and Ying Sheng and Xin Jin and Yanping Huang and Zhifeng Chen and Hao Zhang and Joseph E. Gonzalez and Ion Stoica},
  booktitle = {17th USENIX Symposium on Operating Systems Design and Implementation (OSDI 23)},
  pages     = {663--679},
  year      = {2023},
  publisher = {USENIX Association},
  address   = {Boston, MA},
  month     = jul,
  isbn      = {978-1-939133-34-2},
  url       = {https://www.usenix.org/conference/osdi23/presentation/li-zhouhan}
}

@book{stankovic1998scheduling,
  title     = {Scheduling in Real-Time Systems},
  author    = {Stankovic, John A. and Spuri, Marco and Ramamritham, Krithi and Buttazzo, Giorgio C.},
  publisher = {Springer},
  year      = {1998},
  address   = {Boston, MA, USA}
}

@article{williams2009roofline,
  author  = {Samuel Williams and Andrew Waterman and David A. Patterson},
  title   = {Roofline: An Insightful Visual Performance Model for Multicore Architectures},
  journal = {Communications of the ACM},
  volume  = {52},
  number  = {4},
  pages   = {65--76},
  year    = {2009},
  doi     = {10.1145/1498765.1498785},
  url     = {https://doi.org/10.1145/1498765.1498785}
}

@inproceedings{cao2025moe_lightning,
  author    = {Shiyi Cao and Shu Liu and Tyler Griggs and Peter Schafhalter and Xiaoxuan Liu and Ying Sheng and Joseph E. Gonzalez and Matei Zaharia and Ion Stoica},
  title     = {{MoE-Lightning}: High-Throughput {MoE} Inference on Memory-constrained {GPUs}},
  booktitle = {Proceedings of the 30th ACM International Conference on Architectural Support for Programming Languages and Operating Systems, Volume 1},
  pages     = {715--730},
  publisher = {ACM},
  year      = {2025},
  doi       = {10.1145/3669940.3707267},
  url       = {https://doi.org/10.1145/3669940.3707267}
}

@inproceedings{schick2023toolformer,
  title     = {Toolformer: Language Models Can Teach Themselves to Use Tools},
  author    = {Schick, Timo and Dwivedi-Yu, Jane and Dess{\`i}, Roberto and Raileanu, Roberta and Lomeli, Maria and Hambro, Eric and Zettlemoyer, Luke and Cancedda, Nicola and Scialom, Thomas},
  booktitle = {Advances in Neural Information Processing Systems},
  volume    = {36},
  year      = {2023},
  publisher = {Curran Associates, Inc.}
}

@inproceedings{yao2023tree,
  title     = {Tree of Thoughts: Deliberate Problem Solving with Large Language Models},
  author    = {Yao, Shunyu and Yu, Dian and Zhao, Jeffrey and Shafran, Izhak and Griffiths, Thomas L. and Cao, Yuan and Narasimhan, Karthik},
  booktitle = {Advances in Neural Information Processing Systems},
  volume    = {36},
  year      = {2023},
  publisher = {Curran Associates, Inc.}
}

@inproceedings{madaan2023selfrefine,
  title     = {Self-Refine: Iterative Refinement with Self-Feedback},
  author    = {Madaan, Aman and Tandon, Niket and Gupta, Prakhar and Hallinan, Skyler and Gao, Luyu and Wiegreffe, Sarah and Alon, Uri and Dziri, Nouha and Prabhumoye, Shrimai and Yang, Yiming and Gupta, Shashank and Majumder, Bodhisattwa Prasad and Hermann, Katherine and Welleck, Sean and Yazdanbakhsh, Amir and Clark, Peter},
  booktitle = {Advances in Neural Information Processing Systems},
  volume    = {36},
  year      = {2023},
  publisher = {Curran Associates, Inc.}
}

@inproceedings{besta2024graph,
  title     = {Graph of Thoughts: Solving Elaborate Problems with Large Language Models},
  author    = {Besta, Maciej and Blach, Nils and Kubicek, Ales and Gerstenberger, Robert and Podstawski, Michal and Gianinazzi, Lukas and Gajda, Joanna and Lehmann, Tomasz and Niewiadomski, Hubert and Nyczyk, Piotr and Hoefler, Torsten},
  booktitle = {Proceedings of the AAAI Conference on Artificial Intelligence},
  volume    = {38},
  number    = {16},
  year      = {2024},
  publisher = {AAAI Press}
}

@inproceedings{azar1994balanced,
  title     = {Balanced Allocations},
  author    = {Azar, Yossi and Broder, Andrei Z. and Karlin, Anna R. and Upfal, Eli},
  booktitle = {Proceedings of the Twenty-Sixth Annual ACM Symposium on Theory of Computing},
  pages     = {593--602},
  year      = {1994},
  publisher = {Association for Computing Machinery},
  address   = {New York, NY, USA},
  doi       = {10.1145/195058.195412}
}

@misc{the_state_of_AI_in_2025,
  title        = {The State of {AI} in 2025},
  author       = {{McKinsey \& Company}},
  year         = {2025},
  howpublished = {McKinsey Global Survey},
  url          = {https://www.mckinsey.com/capabilities/quantumblack/our-insights/the-state-of-ai}
}

\end{document}